\documentclass{aa}
\newcommand{\gsim}{\;\lower.6ex\hbox{$\sim$}\kern-7.75pt\raise.65ex\hbox{$>$}\;}
\newcommand{\lsim}{\;\lower.6ex\hbox{$\sim$}\kern-7.75pt\raise.65ex\hbox{$<$}\;}

\newcommand{\kms}{km~s$^{-1}$}
\usepackage{graphicx}
\begin{document}
\title{Metal abundances of RR Lyrae stars in the bar of the Large Magellanic 
Cloud\thanks{Based on data 
collected at the
European Southern Observatory, proposal numbers 62.N-0802, 66.A-0485,
and 68.D-0466} }
%\subtitle{}

\author{R.G. Gratton\inst{1}, 
A. Bragaglia\inst{2},
G. Clementini\inst{2},
E. Carretta\inst{1,2}, 
L. Di Fabrizio\inst{2,3},
M. Maio\inst{2},
and E. Taribello\inst{4}}

\offprints{R.G. Gratton}
 
\institute{INAF-Osservatorio Astronomico di Padova, Vicolo 
 dell'Osservatorio 5, 35122 Padova, ITALY
 %gratton@pd.astro.it
 \and
 INAF-Osservatorio Astronomico di Bologna, Via Ranzani 1, 
     40127 Bologna, ITALY
 \and
 INAF-Centro Galileo Galilei \& Telescopio Nazionale Galileo, 
     PO Box 565, 38700 S.Cruz de La Palma, Spain
 \and
 Dipartimento di Astronomia, Universit\`a di Bologna, 
     Via Ranzani 1, 40127 Bologna, ITALY}

\date{Received 10 December 2003; accepted  4 March 2004}

\abstract{
Metallicities ([Fe/H]) from low resolution spectroscopy obtained with the Very
Large Telescope (VLT) are presented for 98 RR Lyrae and 3 short period Cepheids
in the bar of the Large Magellanic Cloud. Our metal abundances have typical
errors of $\pm 0.17$~dex. The average metallicity of the RR Lyrae stars is
[Fe/H]=$-1.48\pm 0.03\pm 0.06$\ on the scale of Harris (1996). The star-to-star
scatter (0.29~dex) is larger than the observational errors, indicating a real
spread in metal abundances. The derived metallicities cover the range
$-2.12<$[Fe/H]$<-0.27$, but there are only a few stars having [Fe/H]$>-1$. For
the $ab-$type variables we compared our spectroscopic abundances with those
obtained from the Fourier decomposition of the light curves. We find good
agreement between the two techniques, once the systematic offset of 0.2~dex
between the metallicity scales used in the two methods is taken into account.
The spectroscopic metallicities were combined with the dereddened apparent
magnitudes of the variables to derive the slope of the luminosity-metallicity
relation for the LMC RR Lyrae stars: the resulting value is $0.214\pm
0.047$~mag/dex. Finally, the 3 short period Cepheids have [Fe/H] values in the
range $-2.0<$[Fe/H]$<-1.5$. They are more metal-poor than typical LMC RR Lyrae
stars, thus they are more likely to be Anomalous Cepheids rather than the
short period Classical Cepheids that are being found in a number of dwarf
Irregular galaxies.
}

\authorrunning{Gratton R.G. et al.}
\titlerunning{Metal abundance of RR Lyrae stars in the LMC}

   \maketitle

%
%________________________________________________________________

\section{Introduction}

The Large Magellanic Cloud (LMC) is a first step in our knowledge of the
extragalactic astronomical distance scale. A variety of distance estimates to
the LMC have been published in recent years (for references, see Clementini et
al. 2003a). Among those based on Population II indicators, most are tied to
the value of the absolute visual magnitude of the RR Lyrae variables, that is
the radial pulsators within the classic instability strip on the core
He-burning phase of the metal-poor stars (the so-called horizontal branch,
HB). The mean intrinsic luminosity of the RR Lyrae stars depends on
metallicity according to a relation often assumed to be linear:
$M_V(RR)=\alpha {\rm [Fe/H]} + \beta$ (Sandage 1981a,b), and, at fixed metal
abundance, it is also slightly dependent on evolution off the Zero Age
Horizontal Branch (ZAHB; Sandage 1990).

The precise form of the $M_V(RR)-$[Fe/H] relationship is still matter of
debate; and while growing consensus is being reached on its zero point ($\beta
\sim 0.6$ mag at [Fe/H]=$-1.5$, Cacciari \& Clementini 2003, and references
therein), there is still disagreement on the value of the slope since
literature values range from 0.30 to 0.18-0.20 mag/dex (Cacciari 1999,
Carretta et al. 2000). It has even been suggested that $\alpha$ may not be
unique on the metallicity range spanned by the Galactic globular clusters
(GGCs) since two different linear relationships would be followed by the
metal-poor ([Fe/H]$< -1.5$) and the metal-rich ([Fe/H]$> -1.5$) Galactic and
LMC cluster RR Lyrae stars (Rey et al. 2000, Caputo et al. 2000).

Sandage (1993a) found a rather steep slope of the $M_V(RR)-$[Fe/H]
relationship $\alpha \sim 0.30$ mag/dex from the from the so-called
period-shift analysis of the RR Lyrae stars in a  number of Galactic globular
clusters.

The theoretical evolutionary and pulsation models of horizontal branch (HB)
stars favour instead a shallower slope $\alpha \sim 0.2$~mag/dex, and also
provide evidence for a non-linearity that makes the relation steeper for more
metal rich stars, as recently reviewed and summarized by Cacciari \&
Clementini (2003). A mild slope was also found from the Baade-Wesselink method
applied to Galactic field RR Lyrae stars (Fernley et al. 1998), and from the
determination of the magnitude of the HB in globular clusters of M31 (Rich et
al. 2001), However both these studies do not find clear evidence for a break
at [Fe/H]=$-1.5$,  thus raising concerns on the actual universality of the HB
luminosity-metallicity relationship.

The slope of the luminosity-metallicity relation of the RR Lyrae stars can
provide clues on the Galactic globular clusters relative ages and on the time
scale of the Galactic halo formation and early evolution. In fact, given the
almost constant difference in luminosity between Turn-Off and HB stars:
$\Delta{\rm V}^{\rm HB}_{\rm TO}$, found for the GGCs, if $\alpha$ is as steep
as $\sim 0.3$ mag/dex, globular clusters of different metallicities would be
coeval; while if it is $\sim 0.2$, the most metal-rich globular clusters like
47 Tuc would be $\sim 2-3$~Gyr younger than the most metal-poor ones (Sandage
1993b, and references therein).

Furthermore, the entire Population II distance scale would be affected if the
$M_V(RR)-$[Fe/H] relation was found to be not universal and depending instead
on the star formation history and the environment conditions of the host
galaxy.

It is thus important to derive the luminosity-metallicity relationship for RR
Lyrae stars in different extragalactic systems, the LMC in particular. Walker
(1992a) presented both photometry and metallicity estimates (based on the
color magnitude diagrams and the properties of the RR Lyrae variables) for the
LMC globular clusters known to host RR Lyraes, which however are only a few
(7) and cover a small range in metal abundance ($-2.3< {\rm [Fe/H]} < -1.7$,
Walker 1992a, and references therein). Furthermore, these clusters are spread
over a wide region of the sky, and some of them may well be at distances
somewhat different from the average of the LMC: given the small numbers
involved, uncertainties related to the depth of the LMC may be as large as
about 0.1~mag. More recently, several thousands field RR Lyrae variables have
been identified in the Clouds by surveys devoted to the detection of
microlensing events (MACHO: Alcock et al. 1996; OGLE: Udalski et al. 1997).
However, metallicity estimates are available so far only for a handful of them
(Alcock et al. 1996). If we combine the depth uncertainties with possible
evolution off the ZAHB, at least 100 variables are needed to obtain an
accuracy of 0.05 mag/dex in the slope of the luminosity-metallicity relation.

In recent years we began a study of the RR Lyrae variables in two fields
close to the bar of the LMC. Our initial aim was to derive the
mass-metallicity relation for the double mode pulsators (RRd), described in an
earlier paper (Bragaglia et al. 2001). Accurate photometry for these fields in
the Johnson $BV$\ and Cousins $I$\ bands was obtained using the Danish 1.5~m
telescope, and spectroscopy of the RRd variables was obtained using the EFOSC
spectrograph at the 3.6~m ESO telescope. The analysis of the photometric data 
is described in separate papers (Clementini et al. 2003a, Di Fabrizio et al.
2004). Adding spectroscopic observations, these data are well suited to a more
extensive study of the properties of RR Lyrae stars in the LMC, in particular 
for determining their luminosity-metallicity relation, filling the gap left by
previous studies.

We were able to obtain adequate spectroscopic data for about 100 LMC RR Lyrae
stars using FORS1 at VLT, and derived abundances using a variance of the
Preston $\Delta~S$~ method (Preston 1959). An astrophysical discussion of the
$M_V(RR)-$[Fe/H] relationship we derived, and of its impact on the distance to
the LMC, has been already presented (Clementini et al. 2003a). Here, we give
details on the acquisition and reduction of the spectroscopic data (Section
2), on the extraction of line indices (Section 3), and the derivation of the
metallicities (Section 4). We then list the metal abundance of the individual
variables, estimate the associated errors, discuss the properties of the
sample (Section 5), compare the metal abundances of the {\it ab}-type RR Lyrae
with those derived from analysis of the Fourier terms of the light curves
(Section 6), and describe the derivation of the luminosity-metallicity
relation (Section 7). Finally, in Section 8 we comment on the metallicities of
the short period Cepheids present in our sample.

\section{Observations and data reduction}

Spectroscopic data for the LMC targets and for HB stars in a number of
calibrating globular clusters were collected with FORS1 (FOcal Reducer/low
dispersion Spectrograph), mounted at UT1-Antu of the ESO Very Large Telescope
(Paranal, Chile), on UT 21, 22, and 23 December 2001. The first two nights
were rather good (almost photometric, with seeing varying between 0.5\arcsec
~and 1.3\arcsec, but generally less than 1\arcsec), and the last one slightly
worse (light clouds present, and seeing varying between 0.8\arcsec ~and
1.7\arcsec, but generally larger than 1.2\arcsec).

We used FORS1 in multi object (MOS) mode: up to 19 spectra could be obtained in
one exposure, using 19 pairs of movable slitlets, about 20\arcsec ~long, on a
field of view (FoV) of 6.8~arcmin$^2$. We selected the blue grism
GRIS\_600B, covering the 3450 - 5900 \AA ~wavelength range, with a dispersion
of 50 \AA ~mm$^{-1}$ (R $\simeq$ 800), with slits 1\arcsec ~large. In this
mode, each pixel is about 1.2~\AA. Only part of the FoV (about two thirds) was
usable for each pointing, to cover the relevant wavelength range (i.e., $\sim$
3900 - 5100 \AA) cointaining both the Ca{\sc ii} K and the hydrogen Balmer
lines up to H$\beta$.

Pre-imaging frames were required, in conjunction with FIMS (the FORS
Instrumental Mask Simulator package), to prepare all masks (42 in total: 33 on
the LMC, 2 each on M68 and NGC1851, 1 on NGC3201, 4 on $\omega$ Cen). In the
case of the LMC, our original "Field A" and "Field B" (see Clementini et al.
2003a, and Di Fabrizio et al. 
2004), having a side of $\sim 13$~arcmin,
were divided in 4 sub-fields each to fit into the FORS usable FoV. We
were able to fit into the 33 LMC masks about 100 of the 125 RR Lyrae stars, 
and 3 of the 4 short period Cepheids we had
detected  and studied. 

%%%%% place figures here

\begin{figure} 
\caption[]{FORS1 LMC sub-field A1, Nord-East quadrant.
Variables are marked by open circles and identified according to
Di Fabrizio et al. (2004).}
\label{f:fig1a}
\end{figure}

\begin{figure} 
\caption[]{FORS1 LMC sub-field A4, South-East quadrant.
Variables are marked by open circles and identified according to
Di Fabrizio et al. (2004).
}
\label{f:fig1d}
\end{figure} 

\begin{figure} 
\caption[]{FORS1 LMC sub-field A2, Nord-West quadrant.
Variables are marked by open circles and identified according to
Di Fabrizio et al. (2004).
}
\label{f:fig1b}
\end{figure}

\begin{figure} 
\caption[]{FORS1 LMC sub-field A3, South-West quadrant.
Variables are marked by open circles and identified according to
Di Fabrizio et al. (2004).
}
\label{f:fig1c}
\end{figure}

\begin{figure} 
\caption[]{FORS1 LMC sub-field B1, Nord-East quadrant.
Variables are marked by open circles and identified according to
Di Fabrizio et al. (2004).}
\label{f:fig2a}
\end{figure}

\begin{figure} 
\caption[]{FORS1 LMC sub-field B4, South-East quadrant.
Variables are marked by open circles and identified according to
Di Fabrizio et al. (2004).
}
\label{f:fig2d}
\end{figure}

\begin{figure} 
\caption[]{FORS1 LMC sub-field B2, North-West quadrant.
Variables are marked by open circles and identified according to
Di Fabrizio et al. (2004).
}
\label{f:fig2c}
\end{figure}

\begin{figure} 
\caption[]{FORS1 LMC sub-field B3, South-West quadrant.
Variables are marked by open circles and identified according to
Di Fabrizio et al. (2004).
}
\label{f:fig2b}
\end{figure}

Figures 1a-d, 2a-d show the eight LMC FORS1 sub-fields with the variables
identified according to Di Fabrizio et al. (2004) identifiers. Centre of field
coordinates are provided in Table~\ref{t:tab0}.

\begin{table}
\caption{Centre of field coordinates of the eight LMC FORS1 sub-fields}
\begin{tabular}{ccc}
\hline
Field& $\alpha_{2000}$ & $\delta_{2000}$\\
\hline
A1  & 5:23:36.47 & $-$70:34:23\\   
A2  & 5:22:18.47 & $-$70:34:23\\   
A3  & 5:22:18.47 & $-$70:40:43\\   
A4  & 5:23:36.47 & $-$70:40:43\\   
B1  & 5:18:17.97 & $-$71:55:91\\   
B2  & 5:17:19.76 & $-$71:55:91\\   
B3  & 5:17:19.76 & $-$71:66:51\\   
B4  & 5:18:17.97 & $-$71:66:51\\   
\hline
\end{tabular}
\label{t:tab0}
\end{table}

Preparation of the observations was a delicate task, since we wanted to
maximize the number of RR Lyrae's observed at/near minimum light for each
pointing, as this minimizes errors in abundance derivations. We prepared a
special software to help maximizing the efficiency of these observations. The
nights were divided in time slots, considering an overhead of 15 minutes for
each pointing, and an exposure of 30 minutes on the LMC, and of 1-3 minutes on
the GC's. We then used our own positions, periods, and epochs for the LMC RR
Lyrae stars to derive for each time slot the maximum number of targets
observable at minimum light in a single pointing, and filled the remaining
slits with other RR Lyrae variables taken at random phase, and with clump
stars. For the RR Lyrae stars in the calibration clusters M68, NGC\,1851\ and
$\omega$ Cen we used the published light curves (from Walker 1998 for
NGC\,1851; from Clement et al. 1993, and Brocato et al. 1994 for M68; from
Kaluzny et al. 1997 for $\omega$ Cen), and re-derived ephemerides,  to decide
which RR Lyrae star to observe, choosing preferentially those at minimum
light. For the NGC\,3201 variables we used updated ephemerides and finding
charts kindly provided to us in advance of publication by A. Layden and A.
Piersimoni, respectively. The efficiency of our selection is reflected in the
large number of RR Lyrae that could be observed, since a simple division of
the variable density for the available field and acceptable phase would have
given about 50\% less observations than we were actually able to perform. We
obtained adequate spectroscopic data for 101 variables in the LMC, and
observed also 9 RR Lyrae stars, 4 red and 1 blue HB stars in NGC\,1851, 8 RR
Lyrae's and 4 red HB's in NGC\,3201, 13 RR Lyrae's, 7 blue and 1 red HB stars
in M68, and 17 RR Lyrae variables, 1 Anomalous Cepheid (AC) and 1 Population
II Cepheid (P2C) in $\omega$ Cen (see Tables 3, 4, 5 and 6). Multiple
observations were obtained for about 50\% of the LMC variables and for the RR
Lyrae stars in $\omega$ Cen, in particular. Spectra were also obtained for 355
clump stars, their analysis is in progress.

Exposure times were of 60 s for $\omega$ Cen, 120 s for NGC3201, 180 s for M68
and NGC1851, and 1800~s (generally, with a few 1600 and 2100 s exposures) for
the LMC. For the latter, we had to compromise between signal to noise ratio
(S/N) and time resolution, in order to obtain the highest S/N without phase
smearing. The S/N varies, depending on exposure time, sky conditions, airmass
(about 1.4 to 1.7 in all nights), and on actual centering of the star in the
slit, but it is generally of about 35 at 4700~\AA, and about 20 at 3950~\AA, as
estimated by the pixel-to-pixel scatter. These values agree well with
expectations based on the Exposure Time Calculator when typical observing
conditions are considered, thus showing that centering errors did not
introduce significant light losses for most of the stars.

Calibration frames (bias and flat field images, and wavelength calibration
lamps) were obtained during daytime.

\begin{figure} 
\includegraphics[width=8.8cm]{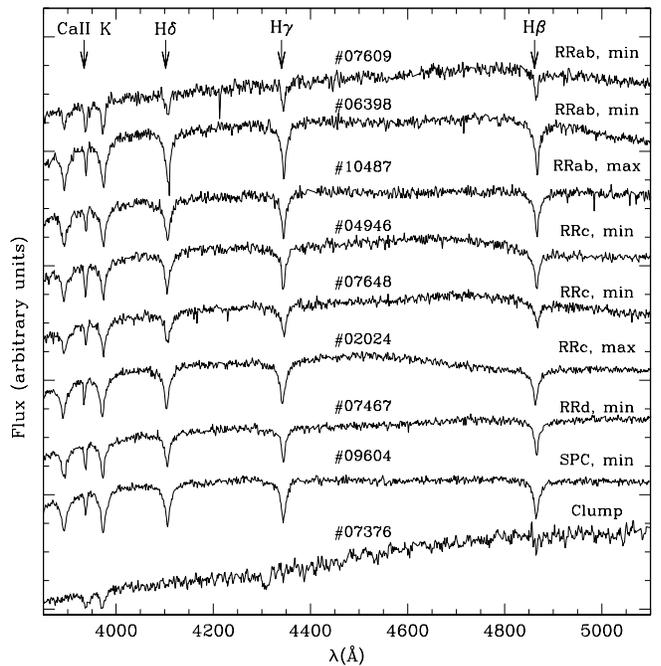}
\caption[]{Examples of spectra of stars in the bar of the LMC
obtained using FORS1. 
The stars are identified according to Di Fabrizio et al.
(2004). The seven upper plots are RR Lyrae variables 
observed at different phases. The two bottom plots are a short
period Cepheid (star \# 9604) and a much cooler clump star (star \# 07376),
respectively. The spectra 
have been offset for clarity, and the main spectral lines
are marked.}
\label{f:fig1}
\end{figure}

The MOS frames were reduced using standard routines in IRAF.\footnote{IRAF is
distributed by the National Optical Astronomy Observatory, which is operated
by the Association of Universities for Research in Astronomy, Inc., under
cooperative agreement with the National Science Foundation.} They were
trimmed, corrected for bias and for the normalized flat field; special care was
devoted to the flat-fielding procedure, since the CCD was read by four
amplifiers, and a "normalization" was needed to eliminate jumps at the
junctions. Up to 19 spectra were present in each frame, and were extracted
with the optimal extraction and automated cleaning options switched on. The
sky contribution was subtracted making use of the slit length. Given the
reasonable seeing conditions, contamination of targets from nearby stars was
reduced to a minimum, except for a few objects. For each science mask, a
HeCdHg lamp was acquired, and used to calibrate in wavelength the spectra,
each one covering a different spectral range, depending on the target
position, as usual in MOS observations, but all comprising the Ca~{\sc ii} K
to H$\beta$ lines. About 10 lines of the calibration lamp were visible for
each aperture, and the resulting dispersion solutions have r.m.s. of about
0.05 \AA. Further cleaning of cosmic rays hits and bad sky subtractions was
done using the clipping option in the $splot$ task; unfortunately a few
spectra were unusable for our procedure because of direct hits on the Ca~{\sc
ii} K line. Figure~\ref{f:fig1} shows examples of the obtained spectra.
Details of the reduction procedures can be found in Taribello (2003).

\begin{figure} 
\includegraphics[width=8.8cm]{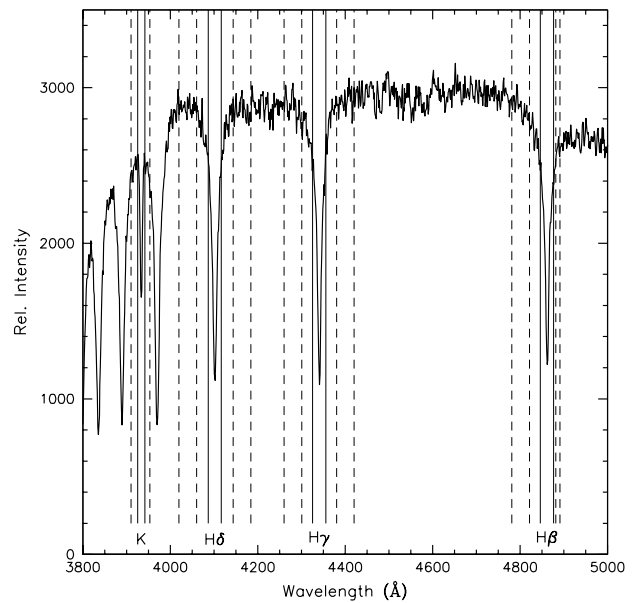}
\caption[]{Example of a spectrum of an RR Lyrae variable. Superposed
are the spectral regions used to define the line indices. The solid lines define
the spectral line bands; dashed lines define the comparison {\it pseudocontinuum}
bands.}
\label{f:fig2}
\end{figure}

\begin{figure} 
\includegraphics[width=8.8cm]{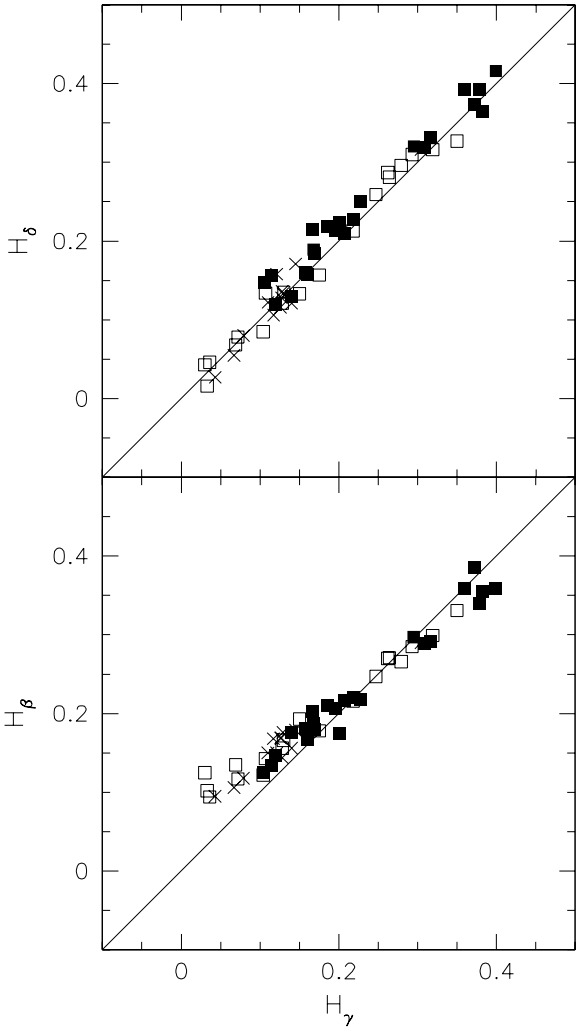}
\caption[]{Correlation between the various indices of the H lines for
the stars in M68 (filled squares), NGC\,1851 (open squares), and
NGC3201 (crosses). Lines represents equality between indices.}
\label{f:fig2b}
\end{figure}

\section{Line indices}

Metal abundances for the RR Lyrae variables can be obtained by comparing the
strength of the Ca~{\sc ii} K line with that of the H lines (Preston 1959).
Preston used spectral types, best suited for photographic spectra; since we
have CCD spectra, we prefer to use line indices.

The spectra were measured using the program ROSA by Gratton (1989). The
following steps were done on each spectrum.

\begin{enumerate}
\item Geocentric radial velocities were measured by cross correlating the
spectra with a suitable template (the RR Lyr variable 2 in M68, observed at
high S/N). The zero point of the radial velocities was obtained by combining
all data for M68, and comparing the resulting average velocity with that
listed in Harris (1996). The error in the radial velocity measured from each
individual spectrum, derived from the square average of the r.m.s. values for
the 48 stars in our sample with multiple observations, is of 40~\kms. This
error includes both the measure uncertainty, which is mainly due to centering
errors of the stars on the slits, and the contribution due to variation of the
radial velocity with phase during the pulsation. The error distribution of the
48 stars with multiple observations is gaussian, with only a few (6) outliers
with errors larger than 70 \kms. The average velocity of these 48 stars,
without applying phase corrections, is 261 $\pm$ 40~\kms (rms scatter), where
25 \kms ~is the internal error (since we have on average 2.4 observations per
star) and 30 \kms ~is the velocity dispersion. These radial velocities are
suited to study the kinematics of the old stellar component in the LMC, which
however is beyond the purposes of the present work and will be discussed in a
separate paper.
\item All spectra were shifted to rest wavelength using the above measured
geocentric radial velocities, and rebinned at costant wavelength step.
\item In order to estimate line indices, instrumental fluxes in a few bands
were measured on the spectra, by simply integrating the spectra within given
limits. These bands were centered on the Ca~{\sc ii} K line, H$\delta$,
H$\gamma$, and H$\beta$. For each line, we also defined two reference
"pseudo-continuum" bands located roughly symmetrically on each side of the
lines. The list of bands we used is given in Table~\ref{t:tab1}; they are
shown in Figure~\ref{f:fig2}.
\item For each line $i$\ we constructed a line index $I_i$, defined as:
\begin{equation}
I_i=1-F_i/[w_{i1}~F(c_{i1})+w_{i2}~F(c_{i2})],
\end{equation}
where $F_i$\ is the instrumental flux measured for the band containing the line,
$F(c_{i1})$\ and $F(c_{i2})$\ are the fluxes measured in the reference continuum
bands, and $w_{i1}$\ and $w_{i2}$\ are weights that take into account the
separation in wavelength between the line and the comparison bands. Note that
$w_{i1}+w_{i2}=1$. By defining the line indices in this way we minimize the
errors in the flux measurements due to the atmospheric dispersion and
centering on the slit, that cause part of the light from the star to fall out
of the slit.
\item The line indices for the three H lines are very well correlated to each
other, except for a small trend of the $H_\beta$\ index, which is slightly larger
for the cooler stars (see Figure~\ref{f:fig2b}). We then defined an index $<H>$\
as simply the average of the three line indices measured for the
hydrogen lines H$_\delta$, H$_\gamma$, and H$_\beta$.
\item Finally, we defined $K$\ the line index for the Ca~{\sc ii} K line.
\end{enumerate}

\begin{table}
\caption{Definition of the spectral bands}
\begin{tabular}{lccc}
\hline
Index & Min. Wavelength & Max. Wavelength & Weight\\
      &    (\AA)        &     (\AA)       & \\
\hline
c$_{11}$   & 3910.0 & 3925.7 & 0.40 \\  
K          & 3925.7 & 3941.7 &      \\  
c$_{12}$   & 3941.7 & 3953.0 & 0.60 \\  
c$_{21}$   & 4020.0 & 4060.0 & 0.50 \\  
H$_\delta$ & 4086.7 & 4116.7 &      \\  
c$_{22}$   & 4144.0 & 4184.0 & 0.50 \\  
c$_{31}$   & 4260.5 & 4300.5 & 0.50 \\  
H$_\gamma$ & 4325.5 & 4355.5 &      \\  
c$_{32}$   & 4380.5 & 4420.5 & 0.50 \\  
c$_{41}$   & 4781.3 & 4821.3 & 0.29 \\  
H$_\beta$  & 4846.3 & 4876.3 &      \\  
c$_{42}$   & 4881.3 & 4891.3 & 0.71 \\  
\hline
\end{tabular}
\label{t:tab1}
\end{table}

We did not apply any correction for the Ca II K interstellar absorption line,
because it is expected to be small for the LMC and the calibrating cluster
stars, that all have reddenings $E(B-V) \la 0.11$ mag (see discussion in
Gratton et al. 1986).

Expected errors $\sigma_i$\ in the indices may be estimated by differentating
eq. (1). If we further assume that the average fluxes within each band are
close to each other, and note that the sum of the weights of the comparison
bands is equal to 1, we may write:
\begin{equation}
\sigma_i~\approx~\sqrt{\Delta}\,[1/\sqrt{\Delta\lambda_i}
                         +w_{i1}/\sqrt{\Delta\lambda_{i1}}
                         +w_{i2}/\sqrt{\Delta\lambda_{i2}}]/(S/N) ,
\end{equation}
where $\Delta$\ is the wavelength step, $\Delta\lambda_i$, $\Delta\lambda_{i1}$,
and $\Delta\lambda_{i2}$\ are the widths of the bands containing the line $i$\
and the {\it continuum} comparison bands, and $w_{i1}$\ and
$w_{i1}$\ are the weights attributed to them.

\begin{figure} 
\includegraphics[width=8.8cm]{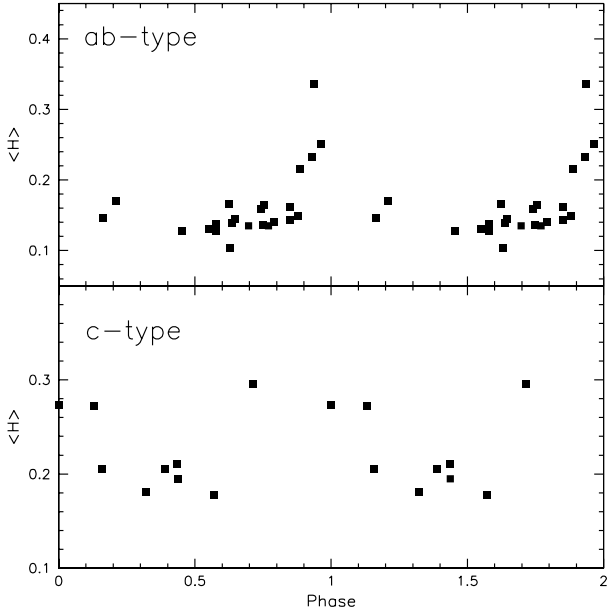}
\caption[]{Run of the $<H>$\ index with phase for stars in the 
calibrating globular clusters. The upper panel is for fundamental mode
pulsators ($ab-$type), and the lower panel for RR Lyrae stars pulsating
in the first overtone ($c-$type).}
\label{f:fig8}
\end{figure}

\begin{figure} 
\includegraphics[width=8.8cm]{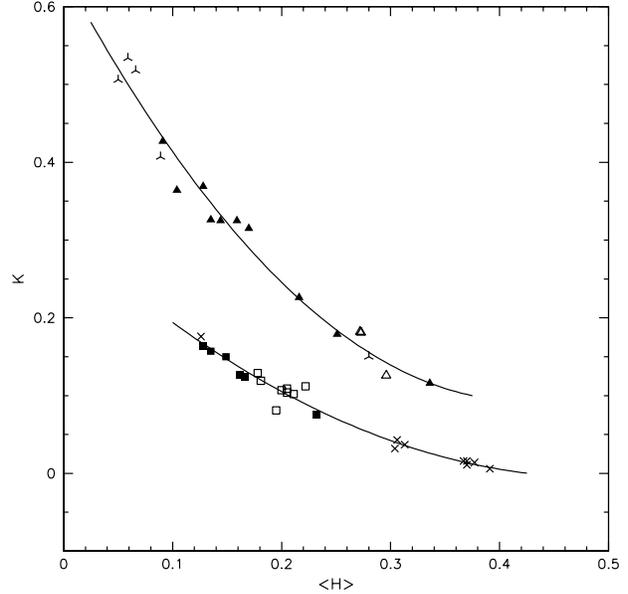}
\caption[]{Relation between the $<H>$\ and $K$\ spectral indices for stars in
the calibration globular clusters: M68 (squares), NGC\,1851 (triangles).
Different symbols are used for $ab-$type RR Lyraes (filled symbols), $c-$type
RR Lyraes (open symbols), and non variable HB stars (crosses with four 
and three
arms, respectively). Superposed are the mean lines for M68 and NGC1851.
A few of the variables shown in the figure have multiple observations 
(see Tables~3 and 4).}
\label{f:fig3}
\end{figure}

\begin{figure} 
\includegraphics[width=8.8cm]{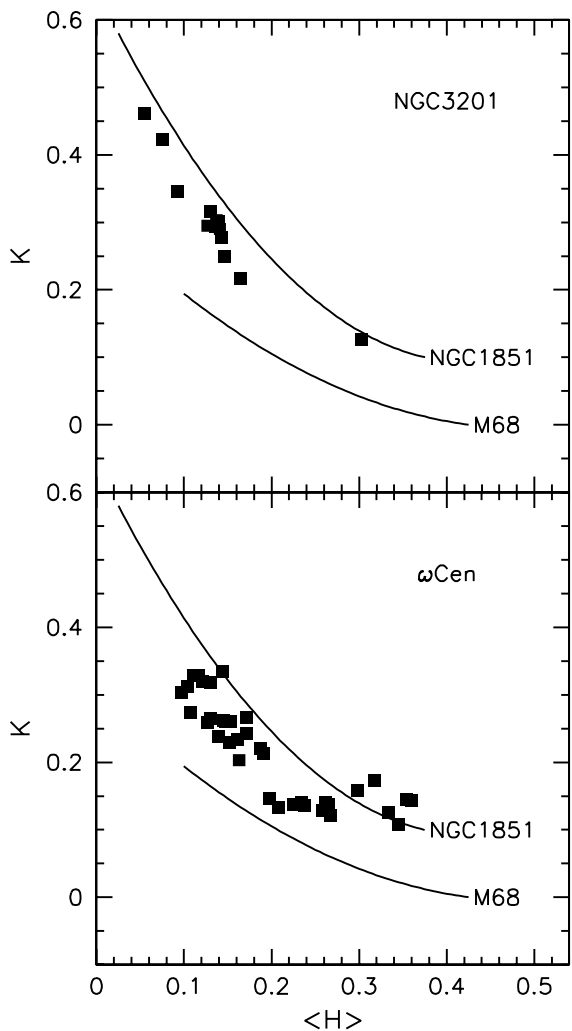}
\caption[]{Relations between the $<H>$\ and $K$\ spectral indices for
variables in NGC3201 (upper panel) and $\omega$~Cen (lower panel).
Superposed are the mean lines for M68 and NGC1851.}
\label{f:fig3b}
\end{figure}

\begin{figure} 
\includegraphics[width=8.8cm]{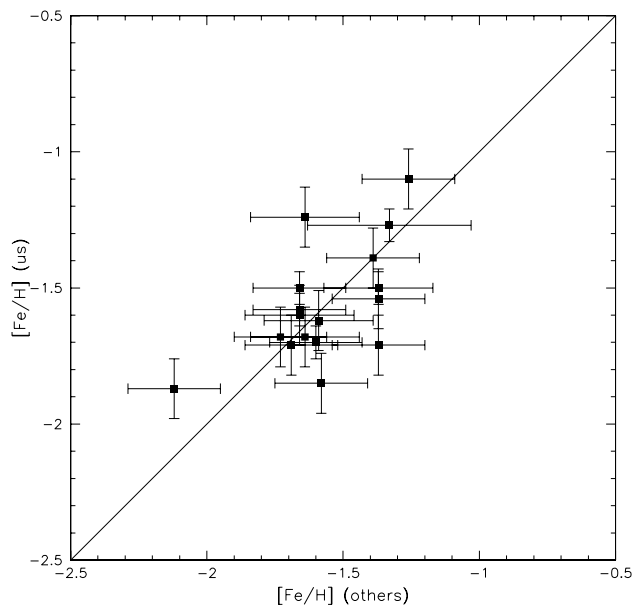}
\caption[]{Comparison between metal abundances from the present analysis and
the average of the values determined by Butler et al. (1978) using the $\Delta
S$\ method, and by Rey et al. (2000) using the $caby$\ photometry, for RR
Lyrae's in the globular cluster $\omega$~Cen. The solid line represents equality
of the two values.}
\label{f:fig4}
\end{figure}

\begin{table*}[htbp]
\begin{center}
\caption {Metallicity indices and metal abundances for the NGC\,1851 stars} 
\begin{footnotesize}
\begin{tabular}{cccccccccc}
\hline
 Star &Type&HJD&   K&  $H_{\beta}$ & $H_{\gamma}$ & $H_{\delta}$ &
$<H>$& M.I.&[Fe/H]\\
\hline
%    1a & &2452265.5278 & 0.126 & 0.316 & 0.304 & 0.290 & 0.303 & 0.884& -1.35\\
~~~1 &ab&2452266.5229 & 0.325 & 0.135 & 0.130 & 0.166 & 0.144 & 0.960& -1.29\\
~~~5 &ab &2452265.5278 & 0.427 & 0.068 & 0.069 & 0.135 & 0.091 & 0.976& -1.28\\
     &   &2452266.5229 & 0.364 & 0.085 & 0.104 & 0.122 & 0.103 & 0.803& -1.42\\
~~~6 &ab&2452266.5229 & 0.315 & 0.157 & 0.175 & 0.178 & 0.170 & 1.162& -1.13\\
~~~7 &ab&2452265.5278 & 0.116 & 0.327 & 0.350 & 0.331 & 0.336 & 0.993& -1.27\\
     &  &2452266.5229 & 0.326 & 0.121 & 0.128 & 0.156  &0.135 & 0.881 &-1.36\\
~~11 &ab&2452265.5278 & 0.179 & 0.259 & 0.247 & 0.247 & 0.251 & 0.964& -1.29\\
     &  &2452266.5229 & 0.369 & 0.134 & 0.107 & 0.143 & 0.128 & 1.043& -1.23\\
~~17 &ab&2452266.5229 & 0.226 & 0.213 & 0.218 & 0.216 & 0.216 & 1.011& -1.25\\
~~21 &c &2452266.5229 & 0.182 & 0.281 & 0.264 & 0.271 & 0.272 & 1.189& -1.11\\
~~22 &ab&2452265.5278 & 0.325 & 0.133 & 0.150 & 0.193 & 0.158 & 1.101& -1.18\\
~~23 &c &2452265.5278 & 0.181 & 0.287 & 0.262 & 0.270 & 0.273 & 1.186& -1.11\\
     &  &2452266.5229 & 0.126 & 0.310 & 0.293 & 0.285 & 0.296 & 0.830& -1.40\\
99999&bHB&2452265.5278 & 0.150 & 0.296 & 0.279 & 0.266 & 0.280 & 0.950& -1.30\\
%    1l & &2452265.5278 & 0.004 & 0.316 & 0.319 & 0.299 & 0.311 &-0.344& -2.33\\
~~129 &rHB&2452266.5229 & 0.518 & 0.043 & 0.030 & 0.125 & 0.066 & 1.132& -1.15\\
~1031 &rHB&2452266.5229 & 0.407 & 0.078 & 0.072 & 0.117 & 0.089 & 0.875 &-1.36\\
~1247 &rHB&2452266.5229 & 0.506 & 0.016 & 0.033 & 0.102 & 0.050 & 0.950& -1.30\\
~1278 &rHB&2452266.5229 & 0.534 & 0.046 & 0.036 & 0.094 & 0.058  &1.125& -1.16\\

\hline
\end{tabular}

Note:									
Star identifiers are from Sawyer-Hogg (1973) for the RR Lyrae stars, and
from Walker (1992b) for the red HB stars (rHB). 
\end{footnotesize}
\end{center}
\label{t:tabl1}
\end{table*}

\begin{table*}[htbp]
\begin{center}
\caption {Metallicity indices and metal abundances for the M68 stars} 
\begin{footnotesize}
\begin{tabular}{ccccccccrc}
\hline
   ~Star &Type & HJD&   K&  $H_{\beta}$ & $H_{\gamma}$ & $H_{\delta}$ &  
$<H>$&M.I.~~&[Fe/H]\\
\hline
   ~~2 &ab &2452264.8400& 0.127 & 0.158 & 0.160 & 0.167 & 0.161 &-0.056& -2.10\\
   ~~8 &c  &2452265.8372 & 0.081 & 0.215 & 0.167 & 0.203 & 0.195& -0.197& -2.22\\
   ~~9 &ab &2452264.8400  & 0.075 & 0.251 & 0.228 & 0.218 & 0.232& -0.060& -2.11\\
       &   &2452265.8372 & 0.124 & 0.160 & 0.158 & 0.181 & 0.166& -0.053& -2.10\\
   ~12 &ab &2452265.8372 & 0.107 & 0.224 & 0.201 & 0.175 & 0.200&  0.010& -2.05\\
   ~13 &c &2452264.8400  & 0.119 & 0.189 & 0.168 & 0.187 & 0.181& -0.005& -2.06\\
   ~17 &ab &2452264.8400  & 0.112 & 0.227 & 0.219 & 0.221 & 0.222&  0.184& -1.91\\
   ~18 &c &2452265.8372 & 0.109 & 0.219 & 0.186 & 0.210 & 0.205&  0.054& -2.02\\
   ~20 &c &2452265.8372 & 0.102 & 0.209 & 0.207 & 0.216 & 0.211&  0.038& -2.03\\
   ~22 &ab &2452265.8372 & 0.150 & 0.130 & 0.140 & 0.176 & 0.149&  0.013& -2.05\\
   ~24 &c  &2452264.8400  & 0.129 & 0.184 & 0.169 & 0.180 & 0.178&  0.044& -2.03\\
   ~25 &ab &2452264.8400  & 0.157 & 0.156 & 0.115 & 0.134 & 0.135& -0.013& -2.07\\
   ~30 &ab &2452264.8400  & 0.164 & 0.119 & 0.119 & 0.147 & 0.128& -0.011& -2.07\\
   ~33 &c  &2452264.8400  & 0.104 & 0.213 & 0.196 & 0.206 & 0.205&  0.021& -2.04\\
   ~86 &bHB &2452264.8400  & 0.014 & 0.373 & 0.372 & 0.385 & 0.377&  0.030& -2.04\\
   112 &bHB &2452264.8400  & 0.016 & 0.392 & 0.359 & 0.359 & 0.370&  0.026& -2.04\\
       &    &2452265.8372 & 0.016 & 0.364 & 0.382 & 0.355 & 0.367&  0.016& -2.05\\
   170 &bHB &2452265.8372 & 0.032 & 0.320 & 0.295 & 0.297 & 0.304& -0.089& -2.13\\
   340 &bHB &2452265.8372  & 0.037 & 0.331 & 0.316 & 0.292 & 0.313&  0.012& -2.05\\
   397 &bHB &2452264.8400  & 0.006 & 0.416 & 0.399 & 0.359 & 0.391& -0.023& -2.08\\
   457 &rHB &2452265.8372 & 0.176 & 0.147 & 0.105 & 0.125 & 0.126&  0.040& -2.03\\
   464 &bHB &2452265.8372 & 0.043 & 0.319 & 0.309 & 0.289 & 0.305&  0.038& -2.03\\
   535 &bHB &2452264.8400  & 0.011 & 0.392 & 0.379 & 0.340 & 0.371& -0.030& -2.08\\
\hline
\end{tabular}
%\end{center}

Note:
Star identifiers are from Sawyer-Hogg (1973) for the RR Lyrae stars, and
from Walker (1994) for the blue and red HB stars (bHB and rHB, respectively).

\end{footnotesize}
\end{center}
\label{t:tabl2}
\end{table*}

\begin{table*}[htbp]
\begin{center}
\caption {Metallicity indices and metal abundances for the NGC\,3201 stars} 
\begin{footnotesize}
\begin{tabular}{cccccccccc}
\hline
 ~Star &Type&HJD&   K&  $H_{\beta}$ & $H_{\gamma}$ & $H_{\delta}$ &  $<H>$&
    M.I.&[Fe/H]\\
\hline
  ~~12 &ab &2452266.8564 &0.216 & 0.171 & 0.145 & 0.179 & 0.165 & 0.506& -1.65\\
  ~~19 &ab &2452266.8564 &0.290 & 0.128 & 0.127 & 0.169 & 0.142 & 0.745& -1.46\\
  ~~20 &ab &2452266.8564 &0.316 & 0.106 & 0.117 & 0.168 & 0.130 & 0.790& -1.43\\
  ~~21 &ab &2452266.8564 &0.303 & 0.116 & 0.126 & 0.169 & 0.137 & 0.782& -1.43\\
  ~~39 &ab &2452266.8564 &0.249 & 0.133 & 0.129 & 0.176 & 0.146 & 0.552& -1.62\\
  ~~40 &ab &2452266.8564 &0.294 & 0.136 & 0.128 & 0.144 & 0.136 & 0.719& -1.48\\
  ~~49 &ab &2452266.8564 &0.278 & 0.158 & 0.121 & 0.150 & 0.143 & 0.689& -1.51\\
  ~~80 &ab &2452266.8564 &0.302 & 0.121 & 0.140 & 0.156 & 0.139 & 0.787& -1.43\\
  ~~10 &rHB &2452266.8564 &0.423 & 0.055 & 0.067 & 0.106 & 0.076 & 0.837& -1.39\\
  ~263 &rHB &2452266.8564 &0.295 & 0.122 & 0.110 & 0.150 & 0.128 & 0.660& -1.53\\
  ~498 &rHB &2452266.8564 &0.461 & 0.027 & 0.043 & 0.095 & 0.055 & 0.818& -1.41\\
  2729 &rHB &2452266.8564 &0.345 & 0.080 & 0.079 & 0.118 & 0.092 & 0.630& -1.56\\
\hline
\end{tabular}

Note:
Star identifiers are from Sawyer-Hogg (1973) for the RR Lyrae stars, and
from Rosenberg et al. (2000, available at http://dipastro.pd.astro.it/globulars/) for the red HB stars (rHB).
\end{footnotesize}
\end{center}
\label{t:tabl3}
\end{table*}

\begin{table*}[htbp]
\caption {Metallicity indices and metal abundances for the variables 
in $\omega$ Cen}
\begin{center}
\begin{scriptsize}
\begin{tabular}{cccccccccccccc}
\hline
 Star &Type&HJD&   K&  $H_{\beta}$ & $H_{\gamma}$ & $H_{\delta}$ &
$<H>$& M.I.&[Fe/H]&$<[Fe/H]>$& [Fe/H] & [Fe/H] & $<$[Fe/H]$>$\\
       &     &	 &    &              &             &               &
     &     &This paper & This paper&  BDE   & $caby$   & Others \\
\hline
 ~71 &c &2452264.8479 & 0.121 & 0.334 & 0.328 & 0.299 & 0.321 & 0.961& -1.29&$-$1.34 &         &       &  \\
  &  &2452264.8518 & 0.130 & 0.324 & 0.325 & 0.306 & 0.318 & 1.041& -1.23&	  &	 &	 &  \\
  &  &2452265.8592 & 0.104 & 0.341 & 0.316 & 0.281 & 0.312 & 0.720& -1.48&	     &      &	    &  \\
 ~72 &P2C&2452264.8479 & 0.354 & 0.135 & 0.129 & 0.172 & 0.145 & 1.137& -1.15&$-$1.27 &$-$1.33&     &$-1.33\pm 0.30$ \\
  & &2452264.8518 & 0.360 & 0.131 & 0.127 & 0.172 & 0.143 & 1.147& -1.14&	 &	&	&   \\
  & &2452265.8592 & 0.298 & 0.160 & 0.142 & 0.171 & 0.158 & 0.933& -1.31&	&      &       &   \\
  & &2452266.8462 & 0.345 & 0.104 & 0.097 & 0.124 & 0.108 & 0.751& -1.46&    &      &	    &	\\
 ~73 &AC &2452264.8479 & 0.187 & 0.231 & 0.211 & 0.222 & 0.221 & 0.756& -1.45&$-$1.50 &$-$1.79&$-$1.60&$-1.66\pm 0.17$\\
  & &2452264.8518 & 0.191 & 0.227 & 0.205 & 0.208 & 0.213 & 0.718& -1.49&	 &	&	&    \\
  & &2452265.8592 & 0.171 & 0.253 & 0.242 & 0.234 & 0.243 & 0.817& -1.41&	&      &       &    \\
  & &2452266.8462 & 0.139 & 0.247 & 0.230 & 0.238 & 0.238 & 0.507& -1.65&	 &	&	&    \\
 ~74 &ab &2452266.8462 & 0.265 & 0.144 & 0.127 & 0.141 & 0.137 & 0.576& -1.60&$-$1.60 &     &$-$1.66&$-1.66\pm 0.20$ \\
 ~98 &c &2452264.8479 & 0.130 & 0.285 & 0.261 & 0.249 & 0.265 & 0.636& -1.55&$-$1.58 &$-$1.65&$-$1.66&$-1.66\pm 0.17$ \\
  & &2452264.8518 & 0.127 & 0.278 & 0.256 & 0.242 & 0.259 & 0.559& -1.61&	 &	&	&    \\
  & &2452265.8592 & 0.097 & 0.326 & 0.296 & 0.289 & 0.303 & 0.591& -1.59&	&      &       &    \\
 ~99 &ab &2452264.8479 & 0.225 & 0.138 & 0.118 & 0.157 & 0.138 & 0.362& -1.77&$-$1.70 &$-$1.29&$-$1.74&$-1.60\pm 0.17$ \\
  & &2452264.8518 & 0.238 & 0.137 & 0.111 & 0.159 & 0.136 & 0.423& -1.72&	 &	&	&    \\
  & &2452265.8592 & 0.262 & 0.126 & 0.136 & 0.159 & 0.140 & 0.583& -1.59&	&      &       &    \\
 100 &ab &2452264.8479 & 0.153 & 0.261 & 0.257 & 0.262 & 0.260 & 0.805& -1.42&$-$1.50 &     &$-$1.37&$-1.37\pm 0.20$ \\
  & &2452264.8518 & 0.148 & 0.262 & 0.258 & 0.259 & 0.260  &0.758& -1.45&	 &	&	&    \\
  & &2452265.8592 & 0.145 & 0.266 & 0.266 & 0.253 & 0.262 & 0.751& -1.46&	&      &       &    \\
  & &2452266.8462 & 0.107 & 0.297 & 0.271 & 0.252 & 0.273 & 0.481& -1.68&	 &	&	&    \\
% 1101 & & & 1.000 & 0.279 & 0.273 & 0.271 & 0.274 & 9.014&  5.15 &-20.7\\
 124 &ab &2452266.8462 & 0.267 & 0.127 & 0.098 & 0.141 & 0.122 & 0.476& -1.68&$-$1.68 &     &$-$1.64&$-1.64\pm 0.20$ \\
 166 &ab &2452265.8490 & 0.234 & 0.133 & 0.132 & 0.158 & 0.141 & 0.435& -1.71&$-$1.71 &$-$0.69&$-$1.67&$-1.37\pm 0.17$ \\
 168 &c &2452265.8490 & 0.161 & 0.259 & 0.217 & 0.225 & 0.234 & 0.653& -1.54&$-$1.54 &$-$1.45&$-$1.33&$-1.37\pm 0.17$ \\
 169 &c &2452265.8490 & 0.171 & 0.271 & 0.266 & 0.261 & 0.266 & 1.023& -1.24&$-$1.24 &      &$-$1.64&$-1.64\pm 0.20$ \\
 170 &ab &2452265.8490 & 0.258 & 0.123 & 0.132 & 0.133 & 0.129 & 0.480& -1.68&$-$1.68 &$-$1.69&$-$1.75&$-1.73\pm 0.17$ \\
%  4101 & & &1 .000 & 0.317 & 0.332 & 0.309 & 0.319 &10.426&  6.28& -25.7\\
 178 &c &2452266.8368 & 0.163 & 0.242 & 0.183 & 0.184 & 0.203 & 0.434& -1.71&$-$1.71 &$-$1.36&$-$1.84&$-1.69\pm 0.17$ \\
 179 &ab &2452266.8368 & 0.197 & 0.139 & 0.136 & 0.164 & 0.146 & 0.265& -1.85&$-$1.85 &$-$1.14&$-$1.78&$-1.58\pm 0.17$ \\
 182 &ab &2452266.8368 & 0.208 & 0.117 & 0.125 & 0.157 & 0.133  &0.246& -1.86&$-$1.87 &$-$2.18&$-$2.09&$-2.12\pm 0.17$ \\
 183 &ab &2452266.8368 & 0.317 & 0.162 & 0.163 & 0.193 & 0.173  &1.200& -1.10&$-$1.10 &$-$0.75&$-$1.49&$-1.26\pm 0.17$ \\
 184 &ab &2452266.8368 & 0.333 & 0.127 & 0.111 & 0.139 & 0.126 & 0.837& -1.39&$-$1.39 &$-$1.06&$-$1.54&$-1.39\pm 0.17$ \\
 186 &c &2452264.8479 & 0.111 & 0.347 & 0.327 & 0.313 & 0.329 & 0.905& -1.34&$-$1.22 &      &	    &		     \\
  & &2452264.8518 & 0.117 & 0.345 & 0.326 & 0.312 & 0.327 & 0.960& -1.29&	 &	&	&    \\
  & &2452265.8592 & 0.144 & 0.355 & 0.318 & 0.331 & 0.335 & 1.305& -1.02&	&      &       &    \\
 187 &c &2452266.8368 & 0.152 & 0.246 & 0.220 & 0.223 & 0.230 & 0.556& -1.62&$-$1.62 &      &$-$1.59&$-1.59\pm 0.20$ \\
\hline
\end{tabular}
%\end{center}

Note:
Star identifiers are from Kaluzny et al. (1997); the type classification for
the Population II Cepheid (P2C) and for the Anomalous Cepheid (AC) is from
Nemec, Nemec, \& Lutz 1994; BDE=Butler et al. (1978); $caby$= Rey et al.
(2000); Others= average of BDE and Rey et al. (1978).
\end{scriptsize}
\end{center}
\label{t:tabl4}
\end{table*}

\section{Metallicity calibration}

$<H>$\ is essentially a measure of the strength of the Hydrogen lines. Within
the temperature range of interest, $<H>$\ is positively correlated with
temperature, and it changes with phase since the temperature of a pulsating
star changes with phase during the pulsation cycle. The run of $<H>$\ with
phase for the globular cluster variables is shown in Figure~\ref{f:fig8}. In
general, $<H>$\ is a function of the temperature $T$, gravity $g$, and metal
abundance [Fe/H], so that $<H>=<H>(T,g,{\rm [Fe/H]})$. $K$\ is a measure of
the strength of the Ca~{\sc ii} K line. In this range of temperatures, $K$\
decreases with temperature. Again, $K$\ is a function of temperature, gravity,
and metal abundance: $K=K(T,g,{\rm [Fe/H]})$. If we now limit ourselves to
stars on the HB\footnote{The three LMC short period Cepheids in our sample were
all treated as being HB stars, we will come back to this point in Section 8.},
we find that for these stars the gravity $g$\ is essentally a function of
temperature and metal abundance. We may then simplify the relations for $<H>$\
and $K$\ and write $<H>=<H>(T,{\rm [Fe/H]})$\ and $K=K(T,{\rm [Fe/H]})$. We
further notice that the dependence of $<H>$\ on metallicity is weak. We then
expect that stars in a cluster (all having the same [Fe/H]) will define a
nearly uniparametric sequences in the $<H>-K$\ plane, with the sequences of
different clusters being shifted according to metallicity. Metallicities (for
horizontal branch stars) can then be derived from the location of the stars in
this plane, provided that suitable calibration sequences are available. This
is the procedure adopted throughout this paper, following the recipe described
below.

First, we plotted the $K$\ index measured on each spectrum of the calibrating
clusters M68 and NGC\,1851  against the $<H>$\ index (see Figure~\ref{f:fig3}
and Table~3 and 4)\footnote{NGC~3201 was not used here because the spectra of
stars in this cluster all have similar $<H>$\ indeces}. Note that
Figure~\ref{f:fig3} contains both variable and constant HB stars. We found
that the RR Lyrae stars have $<H>$\ between 0.08 and 0.33: stars out of this
range are non variables (those with $<H>$\ less than 0.08 are red horizontal
branch stars, mainly observed in NGC1851; while those with $<H>$\ larger than
0.33 are blue horizontal branch stars, mainly observed in M68). RR Lyrae may
be both fundamental mode ($ab$-type) and first overtone  ($c$-type) pulsators;
when observed near minimum, $ab$-type RR Lyrae's have $<H>\sim 0.13$, and
$c$-type ones $<H>\sim 0.19$. However, apart from segregation in the values of
$<H>$\ when observed at minimum light, we found no systematic offset between
the relations defined by variable and non variable stars whithin each cluster.
In practice, we found that for each cluster there is a well defined relation
between $<H>$\ and $K$, with a small scatter around it, although only for
M\,68 and NGC\,1851 the spread in $<H>$\ is large enough to define adequately
the relation over a wide range of values of $<H>$. In fact for NGC\,3201 we do
not have blue HB stars and the variables were observed almost exclusively
around minimum light, while, due to the spread in metallicity the $\omega$ Cen
variables the relation between $<H>$\ and $K$\ for this cluster is very
scattered. On the other hand, since the variables in M68 and NGC\,1851 were
observed when the stars where at different phases and may have different
effective gravities at the same temperature, we expected some thickness in
these relations, since each star is expected to describe a loop in this
diagram during its pulsation cycle. However, Figure~\ref{f:fig3} suggests that
this thickness is small enough that can be neglected in our analysis.

The mean relations drawn from the data in Figure~\ref{f:fig3} are:
\begin{equation}
K_1 = 0.3093 - 1.2815~<H> + 1.3045~H^2
\end{equation}
(valid for $<H>$\ between 0.12 and 0.40) for M68, that has [Fe/H]=$-2.06$\
according to the June 1999 update of Harris (1996) catalogue on
Galactic globular clusters (available at 
http://www.physics.mcmaster.ca/Globular.html), and:
\begin{equation}
K_2 = 0.6432 - 2.6043~<H> + 3.0820~H^2
\end{equation}
(valid for $<H>$\ between 0.04 and 0.34) for NGC 1851, whose metallicity is
[Fe/H]=$-1.26$, again following Harris (1996).

Metallicities for all other stars (both in the LMC and in NGC\,3201 and
$\omega$Cen clusters) were then derived by linear interpolation/extrapolation
between these two relations, entering the measured values of $<H>$\ and $K$\
indices. We defined a metallicity index:
\begin{equation}
M.I.= (K - K_1)/(K_2 - K_1)
\end{equation}
where $K$ is the Ca II K line index of the star and $K_1, K_2$ are derived
entering into equations (3) and (4) the $<H>$ index measured for the star. The
M.I. values derived by this procedure for the HB stars of the 4 calibrating
clusters are listed in Columns 9 and 10 of Tables 3,4,5 and 6, and in Column
11 of Table 7 for the LMC variables. [Fe/H] abundances were then derived from
the relation:
\begin{equation}
{\rm [Fe/H]}= {\rm [Fe/H]}_{\rm M68} + ({\rm [Fe/H]}_{\rm NGC1851} - {\rm
[Fe/H]}_{\rm M68}) {\rm M.I.}
\end{equation}

Of course, these metallicities are less reliable when extrapolations outside
the metallicity range defined by NGC\,1851 and M68 are required. However, as
it will be discussed in Section 5 these extrapolations are necessary only for
a few of the LMC variables in our sample (see Figure 16). Metallicity indices
and corresponding metal abundances obtained for the stars in the calibration
clusters using this procedure are given in Column 10 of Tables~3, 4, 5, and 6.
Metal abundances derived for the LMC variables are provided in Column 12 of
Table~\ref{t:basic} and discussed in Section 5.

As a first test of the accuracy of this abundance calibration, we considered
the case of NGC3201, for which 8 variables were observed. All but one of the
spectra for these stars were taken close to minimum light, i.e. over a rather
small range of values of $<H>$\ (see upper panel of Figure~\ref{f:fig3b} and
Table~5). The average abundance we obtained is [Fe/H]=$-1.49\pm 0.02$, that
nearly coincides with the value of [Fe/H]=$-1.48$\ listed by Harris (1996). The
star-to-star scatter of 0.08~dex is indeed very small, in agreement with the
high S/N of these spectra.

A further test is given by the abundances we obtain from the spectra of
variables in $\omega$~Cen, a cluster that is known to have a large
star-to-star scatter in the [Fe/H] values for individual stars. Butler et al.
(1978) obtained metal abundances for about 50 variables in this cluster, with
average [Fe/H]=$-1.43$\ for the $ab-$type variables (r.m.s. scatter of 0.43
dex), and [Fe/H]=$-1.72$\ for the $c-$type variables (r.m.s. scatter of 0.38
dex). Additional data were obtained by Rey et al. (2000) using the $caby$\
photometry. We obtained a total number of 36 spectra of 17 RR Lyrae stars, 1
Population II Cepheid and 1 Anomalous Cepheid in this cluster (see Table~6 and
lower panel of Figure~\ref{f:fig3b}). For several stars we obtained spectra at
different phases; by comparing the abundances obtained from these spectra, we
obtain a standard error of [Fe/H] determination from a single spectrum of
0.11~dex, similar to the scatter obtained for NGC~3201. These [Fe/H] values
may be compared with the [Fe/H] values obtained by averaging estimates by
Butler et al. (1978) and Rey et al. (2000): these average values were obtained
attributing errors of 0.3 dex to the Butler et al.'s [Fe/H] values, and of 0.2
dex to those of Rey et al., in agreement with the typical errors estimated in
the original papers. This comparison is shown in Figure~\ref{f:fig4}. The
agreement is good: the average difference is $0.01\pm 0.04$~dex, with a
star-to-star scatter of the residuals of 0.18 dex, in good agreement with the
error bars both in our and other determinations.

\begin{figure} 
\includegraphics[width=8.8cm]{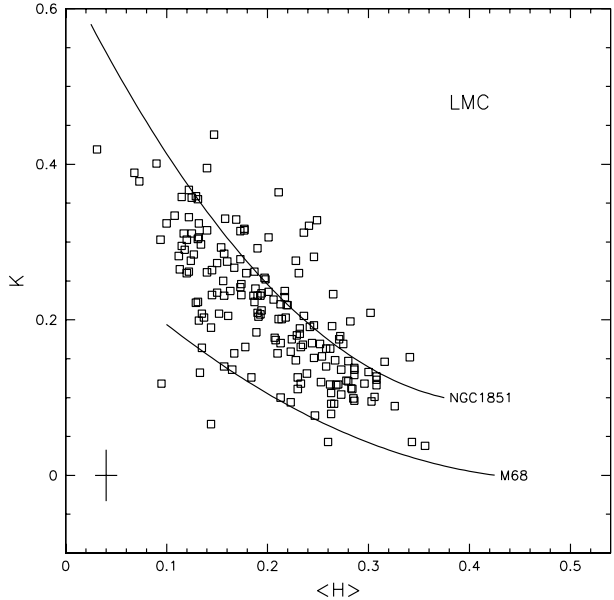}
\caption[]{Relations between the $<H>$\ and $K$\ spectral indices for
variables in the LMC. Superposed are the mean lines for M68 and NGC1851.
The typical error bar is shown at the bottom-left corner.
}
\label{f:fig3c}
\end{figure}

\begin{table*}
\caption{Metallicities for LMC RR Lyrae stars and short period Cepheids}
\scriptsize
\begin{tabular}{rcccccccccccc}
\hline
Star &type   & $V$   & $V_o$ &HJD &$K$&$H_\delta$&$H_\gamma$&$H_\beta$&$<H>$&M.I.&[Fe/H]      &    $<$[Fe/H]$>$    \\
\hline
A1-06332& c & 19.433& 19.101& 2452264.6223 & 0.198 & 0.280 & 0.291 & 0.275 & 0.282 &1.432  & $-0.91\pm 0.27$& $-0.91\pm 0.27$ \\ 
A1-07137& d & 19.413& 19.081& 2452265.5456 & 0.122 & 0.287 & 0.279 & 0.267 & 0.278 &0.655  & $-1.54\pm 0.27$& $-1.54\pm 0.27$ \\
A1-07231& c & 19.322& 18.990& 2452265.6521 & 0.136 & 0.296 & 0.268 & 0.255 & 0.273 &0.749  & $-1.46\pm 0.26$& $-1.46\pm 0.26$ \\
A1-07247&ab & 19.407& 19.075& 2452264.7383 & 0.203 & 0.224 & 0.212 & 0.217 & 0.218 &0.848  & $-1.38\pm 0.21$& $-1.41\pm 0.10$ \\
        &   &       &       & 2452265.5456 & 0.332 & 0.117 & 0.129 & 0.121 & 0.122 &0.806  & $-1.42\pm 0.11$&		      \\
A1-07325&ab & 19.435& 19.103& 2452264.6233 & 0.175 & 0.273 & 0.268 & 0.271 & 0.271 &1.098  & $-1.18\pm 0.26$& $-1.28\pm 0.09$ \\
        &   &       &       & 2452266.7946 & 0.367 & 0.086 & 0.130 & 0.150 & 0.122 &0.983  & $-1.27\pm 0.11$&		      \\
        &   &       &       & 2452265.5456 & 0.148 & 0.295 & 0.251 & 0.255 & 0.267 &0.814  & $-1.41\pm 0.26$&		      \\
A1-07477&ab & 19.183& 18.851& 2452265.5456 & 0.099 & 0.291 & 0.273 & 0.292 & 0.285 &0.481  & $-1.67\pm 0.28$& $-1.67\pm 0.28$ \\
A1-07609&ab & 19.312& 18.980& 2452264.6233 & 0.276 & 0.111 & 0.111 & 0.149 & 0.124 &0.537  & $-1.63\pm 0.11$& $-1.58\pm 0.07$ \\
        &   &       &       & 2452266.7946 & 0.290 & 0.107 & 0.094 & 0.153 & 0.118 &0.564  & $-1.61\pm 0.10$&		      \\
        &   &       &       & 2452265.5456 & 0.234 & 0.204 & 0.171 & 0.208 & 0.194 &0.863  & $-1.37\pm 0.18$&		      \\
A1-07864& c & 19.464& 19.132& 2452264.7383 & 0.189 & 0.270 & 0.214 & 0.213 & 0.232 &0.877  & $-1.36\pm 0.22$& $-1.43\pm 0.18$ \\
        &   &       &       & 2452266.7946 & 0.101 & 0.320 & 0.314 & 0.284 & 0.306 &0.642  & $-1.55\pm 0.30$&		      \\
A1-08094&ab & 19.352& 19.020& 2452265.5456 & 0.222 & 0.107 & 0.125 & 0.153 & 0.129 &0.291  & $-1.83\pm 0.12$& $-1.83\pm 0.12$ \\
A1-08654& d & 19.269& 18.937& 2452264.7383 & 0.201 & 0.223 & 0.210 & 0.199 & 0.211 &0.771  & $-1.44\pm 0.20$& $-1.36\pm 0.14$ \\
        &   &       &       & 2452265.5456 & 0.219 & 0.214 & 0.237 & 0.207 & 0.219 &0.991  & $-1.27\pm 0.21$&		      \\
A1-08720&ab & 19.129& 18.797& 2452264.6233 & 0.043 & 0.340 & 0.369 & 0.320 & 0.343 &0.224  & $-1.88\pm 0.34$& $-1.76\pm 0.20$ \\
        &   &       &       & 2452264.6806 & 0.120 & 0.267 & 0.241 & 0.251 & 0.253 &0.454  & $-1.70\pm 0.24$&		      \\
A1-08788&ab & 19.444& 19.112& 2452265.6521 & 0.159 & 0.230 & 0.226 & 0.214 & 0.223 &0.558  & $-1.61\pm 0.21$& $-1.55\pm 0.14$ \\
        &   &       &       & 2452266.7946 & 0.212 & 0.203 & 0.174 & 0.204 & 0.194 &0.705  & $-1.50\pm 0.18$&		      \\
A1-08837& c & 19.566& 19.234& 2452265.5456 & 0.165 & 0.233 & 0.268 & 0.197 & 0.233 &0.680  & $-1.52\pm 0.22$& $-1.52\pm 0.22$ \\
A1-09154&ab & 19.552& 19.220& 2452265.5456 & 0.235 & 0.157 & 0.135 & 0.157 & 0.150 &0.503  & $-1.66\pm 0.14$& $-1.66\pm 0.14$ \\
A1-09245&ab &\multicolumn{2}{c}{Blazhko}& 2452264.6806 & 0.262 & 0.181 & 0.190 & 0.191 & 0.187 & 0.986 &$-1.27\pm 0.18$& $-1.42\pm 0.10$ \\
        &   &       &       & 2452264.7383 & 0.252 & 0.220 & 0.190 & 0.184 & 0.198 &1.023 & $-1.24\pm 0.19$&		     \\
        &   &       &       & 2452265.5456 & 0.231 & 0.160 & 0.140 & 0.170 & 0.157 &0.532 & $-1.63\pm 0.14$&		     \\
A1-09494&ab & 19.217& 18.885& 2452264.6806 & 0.096 & 0.306 & 0.286 & 0.267 & 0.286 &0.467 & $-1.69\pm 0.28$& $-1.69\pm 0.28$ \\
A1-10113& c & 19.486& 19.154& 2452266.7946 & 0.140 & 0.251 & 0.260 & 0.262 & 0.258 &0.671 & $-1.52\pm 0.25$& $-1.52\pm 0.25$ \\
A1-10360& c &\multicolumn{2}{c}{Shift}& 2452266.7946 & 0.328 & 0.260 & 0.259 & 0.229 & 0.249 &2.243 &$-0.27\pm 0.24$& $-0.27\pm 0.24$ \\
A1-19450&ab & 19.662& 19.330& 2452264.7383 & 0.438 & 0.134 & 0.133 & 0.173 & 0.147 &1.625 & $-0.76\pm 0.13$& $-0.90\pm 0.09$ \\
        &   &       &       & 2452266.7946 & 0.395 & 0.111 & 0.160 & 0.148 & 0.140 &1.300 & $-1.02\pm 0.13$&		     \\
A1-25362&ab & 19.443& 19.111& 2452264.6806 & 0.267 & 0.166 & 0.177 & 0.159 & 0.167 &0.836 & $-1.39\pm 0.15$& $-1.49\pm 0.10$ \\
        &   &       &       & 2452264.7383 & 0.264 & 0.127 & 0.135 & 0.173 & 0.145 &0.629 & $-1.56\pm 0.13$&		     \\
A1-26933&ab & 19.294& 18.962& 2452264.6806 & 0.260 & 0.181 & 0.162 & 0.194 & 0.179 &0.899 & $-1.34\pm 0.17$& $-1.34\pm 0.17$ \\
A1-28066&ab &\multicolumn{2}{c}{Shift}& 2452265.6521 & 0.231 & 0.202 & 0.170 & 0.186 & 0.186 &0.772 &$-1.44\pm 0.17$& $-1.44\pm 0.17$ \\
A2-06398&ab & 19.317& 18.985& 2452265.6300 & 0.116 & 0.329 & 0.304 & 0.293 & 0.308 &0.823 & $-1.40\pm 0.30$& $-1.40\pm 0.30$ \\
A2-06426&ab & 19.185& 18.853& 2452266.6278 & 0.324 & 0.090 & 0.080 & 0.129 & 0.100 &0.589 & $-1.59\pm 0.09$& $-1.59\pm 0.09$ \\
A2-07211&ab &\multicolumn{2}{c}{Incomplete}& 2452266.5441 & 0.226 & 0.198 & 0.207 & 0.213 & 0.206 & 0.912 & $-1.33\pm 0.19$& $-1.33\pm 0.19$ \\
A2-07468&ab & 19.615& 19.283& 2452265.6300 & 0.303 & 0.121 & 0.111 & 0.127 & 0.120 &0.640 & $-1.55\pm 0.11$& $-1.41\pm 0.09$ \\
        &   &       &       & 2452266.5441 & 0.315 & 0.156 & 0.201 & 0.173 & 0.177 &1.227 & $-1.08\pm 0.16$&		     \\
A2-07734&ab &\multicolumn{2}{c}{Shift}& 2452264.6508 & 0.275 & 0.128 & 0.156 & 0.197 & 0.160 & 0.819 &$-1.40\pm 0.15$& $-1.39\pm 0.09$ \\
        &   &       &       & 2452266.5441 & 0.324 & 0.060 & 0.154 & 0.181 & 0.132 & 0.845  & $-1.38\pm 0.12$&  	       \\
A2-08622& c & 19.542& 19.210& 2452265.6300 & 0.129 & 0.308 & 0.268 & 0.281 & 0.286 & 0.780  & $-1.44\pm 0.28$& $-1.60\pm 0.14$ \\
        &   &       &       & 2452266.6278 & 0.148 & 0.241 & 0.217 & 0.225 & 0.228 & 0.502  & $-1.66\pm 0.22$&  	       \\
        &   &       &       & 2452266.5441 & 0.116 & 0.300 & 0.265 & 0.239 & 0.268 & 0.522  & $-1.64\pm 0.26$&  	       \\
A2-08812& c & 19.397& 19.065& 2452265.6300 & 0.193 & 0.265 & 0.250 & 0.223 & 0.246 & 1.037  & $-1.23\pm 0.24$& $-1.38\pm 0.17$ \\
        &   &       &       & 2452266.7118 & 0.151 & 0.257 & 0.242 & 0.240 & 0.246 & 0.676  & $-1.52\pm 0.24$&  	       \\
A2-09604&SPC& 18.932& 18.572& 2452264.6508 & 0.106 & 0.266 & 0.260 & 0.261 & 0.263 & 0.399  & $-1.74\pm 0.25$& $-1.96\pm 0.16$ \\
        &   &       &       & 2452265.7941 & 0.043 & 0.266 & 0.265 & 0.249 & 0.260 &-0.196  & $-2.22\pm 0.25$&  	       \\
        &   &       &       & 2452266.5441 & 0.038 & 0.387 & 0.377 & 0.305 & 0.356 & 0.223  & $-1.88\pm 0.35$&  	       \\
\hline
\end{tabular}
\normalsize
\label{t:basic}
\end{table*}

\addtocounter{table}{-1}

\begin{table*}
\caption{- continued -}
\scriptsize
\begin{tabular}{rcccccccccccc}
\hline
Star &Type & $V$   & $V_o$ &HJD &$K$&$H_\delta$&$H_\gamma$&$H_\beta$&$<H>$&
M.I.&[Fe/H]      &    $<$[Fe/H]$>$    \\

\hline
A2-10214&ab & 19.203& 18.871& 2452265.7941 & 0.297 & 0.119 & 0.130 & 0.152 & 0.134 &0.720 & $-1.48\pm 0.12$& $-1.48\pm 0.12$ \\
A2-10320&SPC& 18.655& 18.295& 2452265.7941 & 0.282 & 0.134 & 0.075 & 0.127 & 0.112 &0.482 & $-1.67\pm 0.10$& $-1.66\pm 0.09$ \\
        &   &       &       & 2452266.7392 & 0.170 & 0.233 & 0.199 & 0.208 & 0.213 &0.562 & $-1.61\pm 0.20$&		     \\
A2-10487&ab & 19.569& 19.237& 2452264.6508 & 0.311 & 0.125 & 0.117 & 0.133 & 0.125 &0.719 & $-1.49\pm 0.11$& $-1.52\pm 0.10$ \\
        &   &       &       & 2452265.6300 & 0.117 & 0.280 & 0.248 & 0.257 & 0.262 &0.494 & $-1.66\pm 0.25$&		     \\
A2-25301&ab & 19.766& 19.434& 2452266.7118 & 0.111 & 0.305 & 0.272 & 0.276 & 0.284 &0.594 & $-1.58\pm 0.27$& $-1.40\pm 0.18$ \\
        &   &       &       & 2452266.7392 & 0.191 & 0.226 & 0.232 & 0.267 & 0.242 &0.974 & $-1.28\pm 0.23$&		     \\
A2-25510&ab & 19.150& 18.818& 2452265.7941 & 0.260 & 0.097 & 0.126 & 0.136 & 0.120 &0.424 & $-1.72\pm 0.11$& $-1.63\pm 0.09$ \\
        &   &       &       & 2452266.5441 & 0.240 & 0.210 & 0.178 & 0.175 & 0.188 &0.848 & $-1.38\pm 0.18$&		     \\
A2-26525&ab & 19.473& 19.141& 2452266.5441 & 0.182 & 0.252 & 0.219 & 0.226 & 0.232 &0.816 & $-1.41\pm 0.22$& $-1.63\pm 0.12$ \\
        &   &       &       & 2452266.6278 & 0.205 & 0.194 & 0.140 & 0.150 & 0.161 &0.409 & $-1.73\pm 0.15$&		     \\
A2-26715& c & 19.378& 19.046& 2452265.7941 & 0.231 & 0.206 & 0.193 & 0.182 & 0.193 &0.836 & $-1.39\pm 0.18$& $-1.38\pm 0.15$ \\
        &   &       &       & 2452266.7118 & 0.138 & 0.293 & 0.311 & 0.253 & 0.286 &0.874 & $-1.36\pm 0.28$&		     \\
A2-26821&ab & 19.624& 19.292& 2452266.6278 & 0.315 & 0.075 & 0.169 & 0.175 & 0.140 &0.867 & $-1.37\pm 0.13$& $-1.37\pm 0.13$ \\
A2-27697& c & 19.166& 18.834& 2452266.5441 & 0.163 & 0.264 & 0.260 & 0.262 & 0.262 &0.910 & $-1.33\pm 0.25$& $-1.33\pm 0.25$ \\
A2-28293&ab & 19.520& 19.188& 2452266.7392 & 0.265 & 0.153 & 0.073 & 0.113 & 0.113 &0.406 & $-1.74\pm 0.10$& $-1.74\pm 0.10$ \\
A2-28665& c &\multicolumn{2}{c}{Shift}& 2452265.6300 & 0.281 & 0.236 & 0.227 & 0.274 & 0.246 & 1.787 & $-0.63\pm 0.24$& $-0.97\pm 0.16$ \\
        &   &       &       & 2452266.5441 & 0.229 & 0.215 & 0.268 & 0.166 & 0.217 &1.037  & $-1.23\pm 0.21$&		      \\
A3-02119& c & 19.659& 19.327& 2452264.7664 & 0.233 & 0.269 & 0.269 & 0.256 & 0.265 &1.585  & $-0.79\pm 0.26$& $-0.79\pm 0.26$ \\
A3-03155& d & 19.209& 18.877& 2452264.5960 & 0.079 & 0.262 & 0.264 & 0.263 & 0.263 &0.151  & $-1.94\pm 0.25$& $-2.01\pm 0.13$ \\
        &   &       &       & 2452264.8216 & 0.136 & 0.163 & 0.170 & 0.163 & 0.165 &0.021  & $-2.04\pm 0.15$&		      \\
A3-03948&ab & 19.292& 18.960& 2452264.8216 & 0.306 & 0.105 & 0.130 & 0.161 & 0.132 &0.751  & $-1.46\pm 0.12$& $-1.46\pm 0.12$ \\
A3-04388& c & 19.427& 19.095& 2452264.5960 & 0.147 & 0.302 & 0.252 & 0.286 & 0.280 &0.910  & $-1.33\pm 0.27$& $-1.33\pm 0.27$ \\
A3-04933&ab & 19.103& 18.771& 2452264.8216 & 0.304 & 0.122 & 0.118 & 0.153 & 0.131 &0.731  & $-1.48\pm 0.12$& $-1.48\pm 0.12$ \\
A3-05148&ab &\multicolumn{2}{c}{Blend}& 2452264.5960 & 0.364 & 0.199 & 0.224 & 0.209 & 0.211 & 1.988 &$-0.47\pm 0.20$& $-0.47\pm 0.20$ \\
A3-05589&ab & 19.574& 19.242& 2452264.7664 & 0.261 & 0.134 & 0.133 & 0.154 & 0.140 &0.578  & $-1.60\pm 0.13$& $-1.60\pm 0.13$ \\
A3-15387&ab & 19.612& 19.280& 2452264.8216 & 0.223 & 0.125 & 0.116 & 0.151 & 0.131 &0.306  & $-1.81\pm 0.12$& $-1.81\pm 0.12$ \\
A3-16249&ab & 19.380& 19.048& 2452264.7664 & 0.203 & 0.144 & 0.130 & 0.136 & 0.137 &0.237  & $-1.87\pm 0.12$& $-1.86\pm 0.09$ \\
        &   &       &       & 2452264.8216 & 0.208 & 0.130 & 0.107 & 0.168 & 0.135 &0.257  & $-1.85\pm 0.12$&		      \\
A3-19711&ab & 19.199& 18.867& 2452264.7664 & 0.303 & 0.084 & 0.084 & 0.115 & 0.094 &0.459  & $-1.69\pm 0.08$& $-1.68\pm 0.08$ \\
A4-02024& c & 19.500& 19.168& 2452265.7640 & 0.117 & 0.276 & 0.269 & 0.266 & 0.270 &0.551  & $-1.62\pm 0.26$& $-1.92\pm 0.14$ \\
        &   &       &       & 2452266.6553 & 0.094 & 0.232 & 0.215 & 0.222 & 0.223 &0.040  & $-2.03\pm 0.21$&		      \\
        &   &       &       & 2452266.5722 & 0.077 & 0.247 & 0.255 & 0.240 & 0.247 &0.045  & $-2.02\pm 0.24$&		      \\
A4-02234& c &\multicolumn{2}{c}{Shift}& 2452265.7640 & 0.207 & 0.193 & 0.188 & 0.196 & 0.193 & 0.663  &$-1.53\pm 0.18$& $-1.53\pm 0.18$ \\
A4-02525&ab & 19.340& 19.008& 2452266.6553 & 0.140 & 0.195 & 0.102 & 0.175 & 0.157 &0.002  & $-2.06\pm 0.14$& $-2.06\pm 0.14$ \\
A4-02623& c & 19.368& 19.036& 2452266.5722 & 0.123 & 0.336 & 0.305 & 0.283 & 0.308 &0.895  & $-1.34\pm 0.30$& $-1.34\pm 0.30$ \\
A4-02636& c & 19.595& 19.263& 2452265.6020 & 0.095 & 0.336 & 0.281 & 0.292 & 0.303 &0.563  & $-1.61\pm 0.29$& $-1.61\pm 0.29$ \\
A4-02767&ab & 19.467& 19.135& 2452266.5722 & 0.401 & 0.019 & 0.088 & 0.164 & 0.090 &0.863  & $-1.37\pm 0.08$& $-1.37\pm 0.08$ \\
A4-03061&ab & 19.631& 19.299& 2452266.5722 & 0.355 & 0.113 & 0.096 & 0.184 & 0.131 &1.001  & $-1.26\pm 0.12$& $-1.26\pm 0.12$ \\
A4-04420& d & 19.409& 19.077& 2452265.7640 & 0.174 & 0.199 & 0.201 & 0.225 & 0.208 &0.551  & $-1.62\pm 0.20$& $-1.32\pm 0.13$ \\
        &   &       &       & 2452266.5722 & 0.292 & 0.212 & 0.145 & 0.214 & 0.190 &1.225  & $-1.08\pm 0.18$&		      \\
A4-04974&ab & 19.384& 19.052& 2452265.6020 & 0.358 & 0.101 & 0.089 & 0.154 & 0.115 &0.870  & $-1.36\pm 0.10$& $-1.35\pm 0.09$ \\
        &   &       &       & 2452266.5722 & 0.236 & 0.208 & 0.197 & 0.199 & 0.201 &0.943  & $-1.31\pm 0.19$&		      \\
A4-12896&ab & 19.588& 19.256& 2452265.7640 & 0.311 & 0.121 & 0.082 & 0.149 & 0.117 &0.658  & $-1.53\pm 0.10$& $-1.53\pm 0.10$ \\
A4-18314&ab & 19.410& 19.078& 2452265.6020 & 0.231 & 0.194 & 0.168 & 0.209 & 0.190 &0.805  & $-1.42\pm 0.18$& $-1.71\pm 0.12$ \\
        &   &       &       & 2452266.5722 & 0.157 & 0.161 & 0.146 & 0.194 & 0.167 &0.157  & $-1.93\pm 0.15$&		      \\
B1-04946& c & 19.432& 19.193& 2452265.7069 & 0.192 & 0.273 & 0.250 & 0.269 & 0.264 &1.193  & $-1.11\pm 0.25$& $-1.50\pm 0.14$ \\
        &   &       &       & 2452266.5993 & 0.118 & 0.205 & 0.240 & 0.256 & 0.233 &0.299  & $-1.82\pm 0.22$&		      \\
        &   &       &       & 2452266.6840 & 0.118 & 0.306 & 0.287 & 0.295 & 0.296 &0.754  & $-1.46\pm 0.29$&		      \\
\hline
\end{tabular}
\normalsize
\label{t:basic}
\end{table*}

\addtocounter{table}{-1}

\begin{table*}
\caption{(- continued -)}
\scriptsize
\begin{tabular}{rcccccccccccc}
\hline
Star    & type & $V$ & $V_o$ &HJD &$K$&$H_\delta$&$H_\gamma$&$H_\beta$&$<H>$&M.I.&[Fe/H]      &    $<$[Fe/H]$>$    \\
\hline
B1-05952&SPC& 18.459& 18.192& 2452265.7069 & 0.378 & 0.044 & 0.050 & 0.125 & 0.073 &0.619 & $-1.56\pm 0.06$& $-1.59:\pm 0.03$ \\
        &   &       &       & 2452265.8201 & 0.389 & 0.060 & 0.048 & 0.096 & 0.068 &0.642 & $-1.55\pm 0.05$&		     \\
        &   &       &       & 2452266.5993 & 0.419 & 0.005 &-0.001 & 0.088 & 0.031 &0.502 & $-1.66\pm 0.05$&		     \\
        &   &       &       & 2452266.6840 & 0.209 & 0.207 & 0.170 & 0.193 & 0.190 &0.657 & $-1.53\pm 0.18$&		     \\
B1-06164& c & 19.057& 18.818& 2452265.8201 & 0.111 & 0.234 & 0.217 & 0.241 & 0.230 &0.225 & $-1.88\pm 0.22$& $-1.79\pm 0.15$ \\
        &   &       &       & 2452266.5993 & 0.157 & 0.198 & 0.207 & 0.224 & 0.210 &0.436 & $-1.71\pm 0.20$&		     \\
B1-06957& c & 19.197& 18.958& 2452265.7069 & 0.223 & 0.216 & 0.167 & 0.176 & 0.187 &0.719 & $-1.48\pm 0.18$& $-1.48\pm 0.18$ \\
B1-07063&ab & 19.629& 19.390& 2452265.8201 & 0.273 & 0.168 & 0.131 & 0.150 & 0.150 &0.719 & $-1.49\pm 0.14$& $-1.49\pm 0.14$ \\
B1-07064& c & 19.122& 18.883& 2452266.5993 & 0.100 & 0.207 & 0.198 & 0.234 & 0.213 &0.032 & $-2.03\pm 0.20$& $-1.96\pm 0.16$ \\
        &   &       &       & 2452266.6840 & 0.092 & 0.277 & 0.251 & 0.260 & 0.263 &0.272 & $-1.84\pm 0.25$&		     \\
B1-07467& d & 19.043& 18.804& 2452265.7069 & 0.126 & 0.240 & 0.221 & 0.230 & 0.230 &0.341 & $-1.79\pm 0.22$& $-1.64\pm 0.10$ \\
        &   &       &       & 2452265.8201 & 0.175 & 0.233 & 0.214 & 0.226 & 0.224 &0.696 & $-1.50\pm 0.21$&		     \\
        &   &       &       & 2452266.5993 & 0.165 & 0.192 & 0.147 & 0.194 & 0.178 &0.273 & $-1.84\pm 0.17$&		     \\
        &   &       &       & 2452266.6840 & 0.179 & 0.257 & 0.321 & 0.236 & 0.272 &1.146 & $-1.14\pm 0.26$&		     \\
B1-07620&ab & 19.078& 18.839& 2452265.7069 & 0.164 & 0.149 & 0.126 & 0.129 & 0.135 &0.017 & $-2.05\pm 0.12$& $-2.05\pm 0.12$ \\
B1-22917&ab & 19.426& 19.187& 2452265.8201 & 0.278 & 0.156 & 0.168 & 0.196 & 0.173 &0.957 & $-1.29\pm 0.16$& $-1.34\pm 0.11$ \\
        &   &       &       & 2452266.5993 & 0.285 & 0.121 & 0.177 & 0.174 & 0.157 &0.852 & $-1.38\pm 0.14$&		     \\
B1-23502&ab & 19.385& 19.146& 2452265.7069 & 0.250 & 0.181 & 0.119 & 0.169 & 0.156 &0.641 & $-1.55\pm 0.14$& $-1.43\pm 0.09$ \\
        &   &       &       & 2452266.5993 & 0.204 & 0.181 & 0.185 & 0.206 & 0.191 &0.625 & $-1.56\pm 0.18$&		     \\
        &   &       &       & 2452266.6840 & 0.237 & 0.215 & 0.242 & 0.194 & 0.217 &1.103 & $-1.18\pm 0.21$&		     \\
B1-24089&ab & 19.365& 19.126& 2452265.7069 & 0.242 & 0.152 & 0.165 & 0.203 & 0.173 &0.730 & $-1.48\pm 0.16$& $-1.31\pm 0.09$ \\
        &   &       &       & 2452266.5993 & 0.329 & 0.183 & 0.161 & 0.162 & 0.169 &1.234 & $-1.07\pm 0.16$&		     \\
        &   &       &       & 2452266.6840 & 0.293 & 0.172 & 0.139 & 0.150 & 0.154 &0.866 & $-1.37\pm 0.14$&		     \\
B2-04780&ab & 19.396& 19.157& 2452264.7935 & 0.357 & 0.118 & 0.119 & 0.139 & 0.125 &0.960 & $-1.29\pm 0.11$& $-1.20\pm 0.08$ \\
        &   &       &       & 2452265.6786 & 0.317 & 0.184 & 0.179 & 0.168 & 0.177 &1.245 & $-1.06\pm 0.16$&  	       \\
        &   &       &       & 2452265.7350 & 0.330 & 0.140 & 0.156 & 0.178 & 0.158 & & $-1.16\pm 0.15$& 		\\
B2-04859&ab & 19.249& 19.010& 2452264.7103 & 0.180 & 0.243 & 0.216 & 0.229 & 0.229 & 0.771 & $-1.44\pm 0.22$& $-1.44\pm 0.22$ \\
B2-05902&ab & 19.121& 18.882& 2452264.7103 & 0.132 & 0.112 & 0.131 & 0.158 & 0.133 &$-$0.155 & $-2.18\pm 0.12$& $-2.12\pm 0.11$ \\
        &   &       &       & 2452265.5716 & 0.092 & 0.275 & 0.264 & 0.261 & 0.266 & 0.290 & $-1.83\pm 0.26$&		      \\
B2-06020&ab &\multicolumn{2}{c}{Shift}& 2452264.7103 & 0.208 & 0.187 & 0.123 & 0.145 & 0.152 &0.365&$-1.77\pm 0.14$& $-1.86\pm 0.11$ \\
        &   &       &       & 2452265.5716 & 0.126 & 0.220 & 0.182 & 0.151 & 0.184 &0.057 & $-2.01\pm 0.17$&		     \\
B2-06255& c & 19.264& 19.025& 2452266.7669 & 0.232 & 0.152 & 0.161 & 0.209 & 0.174 &0.672 & $-1.52\pm 0.16$& $-1.52\pm 0.16$ \\
B2-06440&ab & 19.247& 19.008& 2452265.5716 & 0.169 & 0.274 & 0.272 & 0.280 & 0.275 &1.083 & $-1.19\pm 0.27$& $-1.19\pm 0.27$ \\
B2-06470& d & 19.207& 18.968& 2452266.7669 & 0.334 & 0.120 & 0.038 & 0.167 & 0.108 &0.700 & $-1.50\pm 0.09$& $-1.48\pm 0.09$ \\
        &   &       &       & 2452265.5716 & 0.169 & 0.263 & 0.238 & 0.254 & 0.252 &0.880 & $-1.36\pm 0.24$&		     \\
B2-06798&ab & 19.253& 19.014& 2452266.7669 & 0.170 & 0.278 & 0.244 & 0.211 & 0.244 &0.821 & $-1.40\pm 0.23$& $-1.20\pm 0.13$ \\
        &   &       &       & 2452265.5716 & 0.314 & 0.159 & 0.169 & 0.191 & 0.173 &1.184 & $-1.11\pm 0.16$&		     \\
B2-07442&ab & 19.426& 19.187& 2452265.5716 & 0.284 & 0.112 & 0.104 & 0.165 & 0.127 &0.600 & $-1.58\pm 0.11$& $-1.58\pm 0.11$ \\
B2-07648& c & 19.384& 19.145& 2452265.5716 & 0.127 & 0.328 & 0.291 & 0.305 & 0.308 &0.930 & $-1.32\pm 0.30$& $-1.32\pm 0.30$ \\
B2-19037&ab & 19.701& 19.462& 2452265.5716 & 0.359 & 0.123 & 0.070 & 0.193 & 0.129 &0.998 & $-1.26\pm 0.12$& $-1.26\pm 0.12$ \\
B3-01408&ab & 19.342& 19.103& 2452266.8216 & 0.262 & 0.092 & 0.108 & 0.165 & 0.122 &0.450 & $-1.70\pm 0.11$& $-1.70\pm 0.11$ \\
B3-01575&ab & 19.250& 19.011& 2452265.7350 & 0.295 & 0.103 & 0.102 & 0.140 & 0.115 &0.562 & $-1.61\pm 0.10$& $-1.61\pm 0.10$ \\
B3-02055&ab &\multicolumn{2}{c}{Incomplete}& 2452266.8216 & 0.131 & 0.268 & 0.233 & 0.217 & 0.239 &0.450  & $-1.70\pm 0.23$& $-1.70\pm 0.23$ \\
B3-02249&ab & 19.346& 19.107& 2452264.7935 & 0.237 & 0.158 & 0.141 & 0.190 & 0.163 &0.619  & $-1.56\pm 0.15$& $-1.56\pm 0.15$ \\
B3-02517& c & 19.695& 19.456& 2452265.7350 & 0.152 & 0.356 & 0.350 & 0.316 & 0.341 &1.427  & $-0.92\pm 0.33$& $-0.92\pm 0.33$ \\
B3-02884&ab & 19.629& 19.390& 2452265.6786 & 0.190 & 0.138 & 0.150 & 0.144 & 0.144 &0.214  & $-1.89\pm 0.13$& $-1.90\pm 0.09$ \\
        &   &       &       & 2452265.7350 & 0.199 & 0.121 & 0.133 & 0.143 & 0.132 &0.191  & $-1.91\pm 0.12$&		      \\
B3-03347& d & 19.204& 18.965& 2452265.6786 & 0.184 & 0.183 & 0.178 & 0.205 & 0.189 &0.473 & $-1.68\pm 0.18$& $-1.65\pm 0.13$ \\
        &   &       &       & 2452265.7350 & 0.177 & 0.209 & 0.187 & 0.223 & 0.207 &0.563 & $-1.61\pm 0.20$&		     \\
B3-03400&ab & 19.469& 19.230& 2452264.7935 & 0.153 & 0.254 & 0.255 & 0.255 & 0.254 &0.760 & $-1.45\pm 0.24$& $-1.45\pm 0.24$ \\
B3-03625& c &\multicolumn{2}{c}{Shift}& 2452264.7935 & 0.089 & 0.349 & 0.321 & 0.308 & 0.326 &0.639  &$-1.55\pm 0.32$& $-1.48\pm 0.17$ \\
        &   &       &       & 2452265.7350 & 0.133 & 0.313 & 0.293 & 0.294 & 0.300 & 0.935  & $-1.31\pm 0.29$&  	       \\
        &   &       &       & 2452266.8216 & 0.112 & 0.286 & 0.285 & 0.278 & 0.283 & 0.598  & $-1.58\pm 0.27$&  	       \\
B3-04008& c & 19.281& 19.042& 2452265.7350 & 0.136 & 0.303 & 0.286 & 0.268 & 0.286 & 0.860  & $-1.37\pm 0.28$& $-1.37\pm 0.28$ \\
B3-04179& c & 19.173& 18.934& 2452264.7935 & 0.121 & 0.284 & 0.282 & 0.274 & 0.280 & 0.668  & $-1.53\pm 0.27$& $-1.53\pm 0.27$ \\
B3-14449&ab & 19.514& 19.275& 2452265.6786 & 0.232 & 0.124 & 0.146 & 0.165 & 0.145 &0.145  & $-1.70\pm 0.13$& $-1.70\pm 0.13$ \\
B4-01907&ab & 19.286& 19.047& 2452264.5351 & 0.104 & 0.282 & 0.269 & 0.268 & 0.273 &0.273  & $-1.70\pm 0.26$& $-1.52\pm 0.18$ \\
        &   &       &       & 2452264.5656 & 0.163 & 0.275 & 0.241 & 0.257 & 0.258 &0.258  & $-1.36\pm 0.25$&		      \\
B4-02811& c & 19.384& 19.145& 2452264.5351 & 0.205 & 0.245 & 0.219 & 0.243 & 0.236 &0.236  & $-1.23\pm 0.23$& $-1.24\pm 0.14$ \\
        &   &       &       & 2452264.5656 & 0.254 & 0.171 & 0.185 & 0.235 & 0.197 &0.197  & $-1.24\pm 0.19$&		      \\
B4-03033&ab &\multicolumn{2}{c}{Incomplete}& 2452264.5351  & 0.219 & 0.232 & 0.212 &0.215  & 0.220 & 1.000& $-1.26\pm 0.21$& $-1.36\pm 0.15$ \\
        &   &       &       & 2452264.5656 & 0.168 & 0.266 & 0.231 & 0.208 & 0.235 &0.235  & $-1.48\pm 0.22$&		      \\
B4-04749& c & 19.314& 19.075& 2452264.5351 & 0.246 & 0.175 & 0.170 & 0.176 & 0.174 &0.174  & $-1.45\pm 0.16$& $-1.40\pm 0.13$ \\
        &   &       &       & 2452264.5656 & 0.220 & 0.198 & 0.223 & 0.218 & 0.213 &0.213  & $-1.31\pm 0.20$&		      \\
B4-10811&ab & 19.431& 19.192& 2452264.5656 & 0.201 & 0.231 & 0.187 & 0.223 & 0.214 &0.214  & $-1.42\pm 0.20$& $-1.42\pm 0.20$ \\
\hline
\end{tabular}
\normalsize

Note: $V$\ and $V_0$\ are mean apparent and dereddened intensity averaged
magnitudes taken from Di Fabrizio et al. (2004). Shift means that there is an
offset between light curves obtained in 1999 and in 2001; Incomplete means
that the light curve is incomplete, so that no meaningful average magnitudes
could be obtained; Blazhko means that the light curve shows evidence of a
significant Blazhko effect (Blazhko 1907); Blend means that the stellar image
appears blended in some frames; The metal abundance of B1-05952 is uncertain
(see Section 8).
\label{t:basic}
\end{table*}

\section{Metal abundances of the LMC variables}

Figure~\ref{f:fig3c} shows the run of the $K$\ spectral index as a function of
the $<H>$\ for the variables in the two fields of the LMC. Most of the stars
lie between the ridge lines for M68 ([Fe/H]=$-2.06$) and NGC1851 	
([Fe/H]=$-1.26$), with only a few stars above the latter line (i.e. more
metal-rich than this cluster). We obtained metallicity determinations for 101
LMC variables from the analysis of 168 spectra. According to Di Fabrizio et
al. (2004) 98 of these stars are RR Lyrae, and 3 are short period Cepheids.
Table~\ref{t:basic} lists the metal abundances for the stars in the two fields
of the LMC. Column 1 gives the variable identification with the first two
characters indicating the FORS1 sub-field where the star is located (see
Figures 1a-d, 2a-d), and the remaining digits giving Di Fabrizio et al. (2004)
identification number. Column 2 gives the variable type (ab: fundamental mode;
c: first overtone; d: double-mode pulsator; SPC: Short Period Cepheid).
Columns 3 and 4 give the mean observed and dereddened $V$\ magnitudes taken
from Di Fabrizio et al. (2004). According to Clementini et al. (2003a)
reddenings of $E(B-V)$=0.116 and 0.086 mag were adopted for the stars in field
A and B, respectively. Column 5 gives the Heliocentric Julian Day (HJD) of
observation at mid exposure. Columns 6 to 9 give the measured line indices,
and Column 10 gives the value of $<H>$\ (the average for the three H line
indices). Column 11 gives the metallicity index (M.I.) and Column 12 gives the
metallicity obtained for each spectrum from our calibration of the line
indices. Tabulated errors are obtained as detailed below. Finally, Column 13
gives the final value of the metallicity for each variable obtained by a
weighted average of the metallicity determinations from individual spectra.
These metallicities are tied to Harris (1996) metal abundance  for the
Galactic globular clusters NGC 1851, NGC3201, and M68  ([Fe/H]=$-1.26$,
$-1.48$, and $-2.06$, respectively, to compare with $-1.33$, $-1.56$, and
$-2.09$ of Zinn \& West 1984 metallicity scale), that we used as calibrators.
Thus our metallicities are on a metallicity scale that,  on average, is 0.06
dex more metal rich than the Zinn \& West one\footnote{There is no clear
offset between the Harris (1996) and Zinn \& West (1984) metallicity scales,
the mean offset being only $0.02\pm 0.01$~dex (113 clusters), once a few 
clearly discrepant objects are eliminated. We consider the 0.06 dex difference
found for the three calibrating clusters of the present analysis as a possible
zero point error in our abundance scale.}.

We note  that all stars have [Fe/H]$\leq -0.9$, except A3-05148, that is
flagged as a blend by Di Fabrizio et al. (this is the only star with this
label in our sample), A3-02119 and A1-10360, $c-$type variables for which we
obtain [Fe/H]=$-0.79\pm 0.26$ and $-0.27\pm 0.24$, respectively. It should
also be noted that abundances for stars whose metallicity is larger than
[Fe/H]=$-1.26$\ (the metallicity of NGC1851) are actually obtained by
extrapolation, and are then much more uncertain: this applies in particular to
the stars with [Fe/H]$>-1.0$. On the whole we think that there is little
evidence for RR Lyrae stars more metal rich than [Fe/H]$=-1$\ in the LMC, from
our data.

To estimate observational errors in the [Fe/H] values, we have compared results
obtained from multiple observations of the same star. Since the ridge lines
for the two comparison clusters tend to converge at large values of $<H>$\
(i.e., high temperatures, that is spectra taken far from minimum light), we
expect that the accuracy of the metallicities is a function of the phase at
which the spectra were taken. We have then divided the stars into three groups
according to the average strength of the $<H>$\ parameter. For each group we
estimated the quadratic mean  of the r.m.s., weighting values for individual
stars according to the number of observations; results are shown in
Table~\ref{t:tab4}. As expected, the r.m.s. scatter increases with increasing
strength of the H lines. We then made a linear fit throughout these data, and
assumed that the error in the metallicity determination is given by:
\begin{equation}
\sigma{\rm [Fe/H]}=1.027~<H>-0.017.
\end{equation}

These empirical estimates of the errors in our metallicities can be compared
with the values expected from the quality of our spectra (that we remind are
within expectations on the basis of the Exposure Time Calculator). Errors are
mainly due to the $K$\ index, the contribution by errors in the $<H>$\ index
(evaluated to be 0.014 by comparing the values obtained from different lines)
being smaller, although not entirely negligible. By applying eq. (2), the
expected error in the $K$\ index is 0.029 for a typical spectrum in the LMC.
This value may be compared with that obtained from multiple spectra of the
same star taken at very similar phase. Using eight independent pairs of
observations for which the $<H>$\ index differs less than 0.01, the error in
individual $K$\ measurements is 0.033, in good agreement with expectations. If
we now take into account the sensitivity of metallicity to variations of the
$K$\ index (as a function of $<H>$), that can be obtained by combining eqs.
(3) and (4), and the small contribution given by the errors in $<H>$\
(provided by the slopes of the eqs. (3) and (4)), we may estimate the
predicted errors for typical RR Lyrae stars of the LMC: the expected values
for the three bins of Table~\ref{t:tab4} are 0.157, 0.189, and 0.226~dex,
respectively. These values, listed in the last column of Table~\ref{t:tab4},
agree well with the observed errors, confirming that the accuracy of our
abundance determinations is within expectations.

Multiple observations of the same variable were then combined, weighting
individual determinations according to the error estimated in this way thus
producing the average values listed in Column 13 of Table~\ref{t:basic}.

\begin{figure} 
\includegraphics[width=8.8cm]{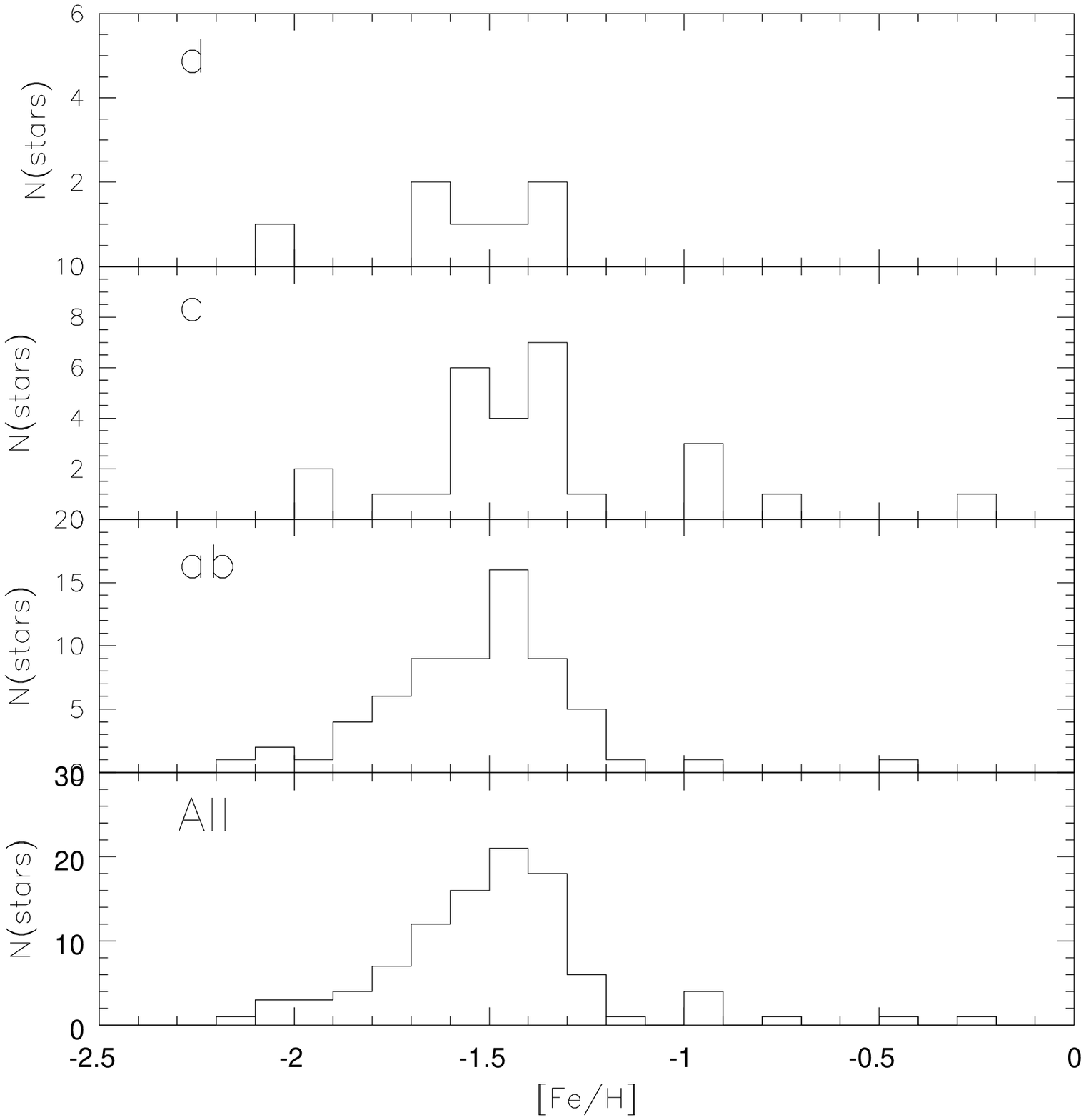}
\caption[]{Metallicity distribution for RR Lyrae stars in the LMC. The bottom panel
shows the distribution of all variables; the other three panels show the
distribution of $ab-$, $c-$\, and $d-$type RR Lyrae stars, respectively.}
\label{f:fig5}
\end{figure}

The metallicity distribution of the RR Lyrae stars in our two fields of the
LMC is given in Table~\ref{t:tab5} and shown in Figure~\ref{f:fig5}. The
average metallicity of the sample is $<$[Fe/H$>=-1.48\pm 0.03\pm 0.06$~dex,
where the last error bar accounts for possible zero point errors in our 
calibration. The
star-to-star scatter (r.m.s=0.29~dex) is clearly larger than the mean
quadratic error of 0.17~dex of the abundances for individual
stars\footnote{This error is smaller than the values quoted in
Table~\ref{t:tab4} because on average about 1.6 observations were obtained for
each star}. By quadratically subtracting the two values, the intrinsic
star-to-star scatter is 0.23~dex. There is no clear systematic offset between
field A and B: the average abundances are $-1.45\pm 0.04$\ and $-1.52\pm
0.04$~dex, respectively. The distribution of Figure~\ref{f:fig5} appears to be
somewhat skewed, with more stars more metal-poor than the average value than
stars more metal rich than this value.

If we divide the variables according to the Bailey types, we find average
metallicities of [Fe/H]=$-1.51\pm 0.03$~dex for $ab-$type RR Lyrae's,
[Fe/H]=$-1.36\pm 0.07$~dex for $c-$type ones, and finally [Fe/H]=$-1.57\pm
0.09$~dex for double pulsators. Differences between values for different types
are quite small; the somewhat larger value found for $c-$type variables may be
due to a few stars with larger error bars, for which we found metallicities
[Fe/H]$>-1$. Note that since on average $c-$type variables have higher
temperatures, they have weaker Ca~{\sc ii} K lines, making the errors of the
metal abundance determinations larger.

\begin{table}
\caption{Errors as a function of $<H>$\ for the LMC stars}
\begin{tabular}{lccccc}
\hline
$<H>$~ bin & n. stars & $<H>$ & r.m.s.([Fe/H]) & Expected \\
\hline
~~~~~$<0.17$  & 16 & 0.151 & 0.138 & 0.157\\
$0.17 - 0.22$ & 18 & 0.203 & 0.193 & 0.189\\
$>0.22$       & 14 & 0.259 & 0.249 & 0.226\\
\hline
\end{tabular}
\label{t:tab4}
\end{table}

\begin{table}
\caption{Metallicity distribution and average magnitudes of the RR Lyrae
stars in the LMC}
\begin{tabular}{lcccc}
\hline
$[$Fe/H$]$~ bin & n. stars & $<[$Fe/H$]>$ & $<V_0>$ \\
\hline
~~~~~$[$Fe/H$]<-2.1$ & ~1 & $-$2.12 & $18.88$         \\
$-2.1<[$Fe/H$]<-2.0$ & ~3 & $-$2.04 & $18.91\pm 0.05$ \\
$-2.0<[$Fe/H$]<-1.9$ & ~3 & $-$1.93 & $19.15\pm 0.15$ \\
$-1.9<[$Fe/H$]<-1.8$ & ~4 & $-$1.84 & $19.12\pm 0.08$ \\
$-1.8<[$Fe/H$]<-1.7$ & ~7 & $-$1.73 & $19.04\pm 0.08$ \\
$-1.7<[$Fe/H$]<-1.6$ & 12 & $-$1.64 & $19.02\pm 0.05$ \\
$-1.6<[$Fe/H$]<-1.5$ & 15 & $-$1.54 & $19.10\pm 0.03$ \\
$-1.5<[$Fe/H$]<-1.4$ & 21 & $-$1.45 & $19.09\pm 0.04$ \\
$-1.4<[$Fe/H$]<-1.3$ & 18 & $-$1.35 & $19.07\pm 0.03$ \\
$-1.3<[$Fe/H$]<-1.2$ & ~6 & $-$1.24 & $19.20\pm 0.07$ \\
$-1.2<[$Fe/H$]<-1.1$ & ~1 & $-$1.19 & $19.01$         \\
$-1.1<[$Fe/H$]<-1.0$ & ~0 &         &                 \\
$-1.0<[$Fe/H$]<-0.9$ & ~4 & $-$0.92 & $19.30\pm 0.10$ \\
$-0.9<[$Fe/H$]<-0.8$ & ~0 &         &                 \\
$-0.8<[$Fe/H$]<-0.7$ & ~1 & $-$0.79 & $19.33$         \\
$-0.7<[$Fe/H$]$      & ~2 & $-$0.37 &                 \\
\hline
\end{tabular}
\label{t:tab5}
\end{table}

\section{Comparison with abundances from light curve Fourier analysis}

Metallicities from the parameters of the Fourier decomposition of the $V$
light curves were derived for 29 RRab's in our photometric sample, 7 of which
do not have spectroscopic metal abundances, applying Jurcsik \& Kov\'acs (1996)
and Kov\'acs \& Walker (2001) techniques (see Di Fabrizio et al. 2004, 
for details).

\begin{figure} 
\includegraphics[width=8.8cm]{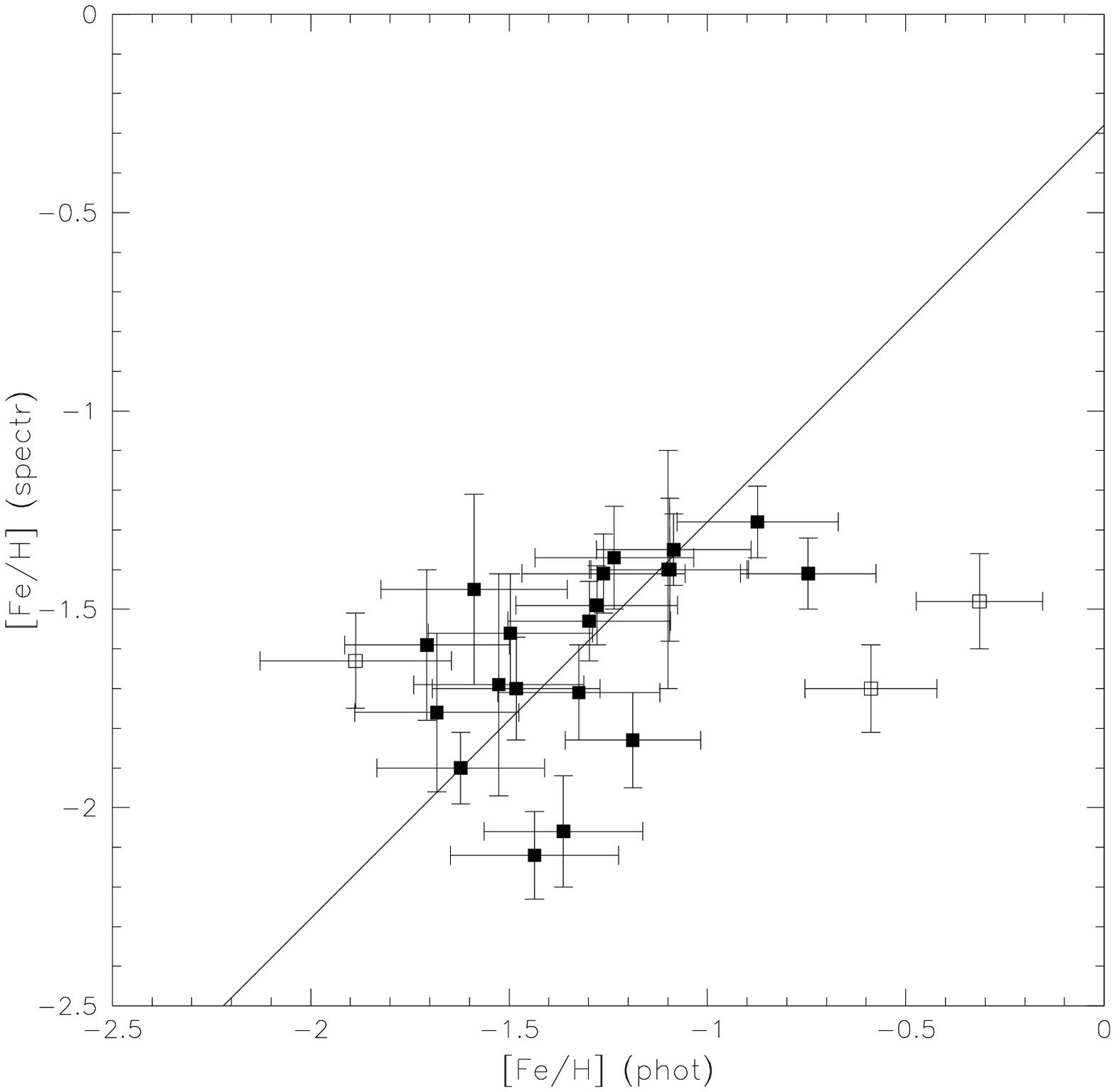}
\caption[]{Comparison between metal abundances for LMC RR Lyrae's obtained
from spectroscopy and from analysis of the terms of a Fourier decomposition
of the light curves following Jurcsik \& Kov\'acs (1996) and
Kov\'acs \& Walker (2001) technique. Open symbols mark stars for 
which discrepant reddening values were derived using Sturch (1966) 
method. The solid line is the mean offset between the two
metallicity sets.}
\label{f:fig5b}
\end{figure}

These metal abundances are on the metallicity scale defined by Jurcsik (1995)
which is, on average, about 0.2 dex more metal rich than Harris (1996) scale.
The average metallicity of this subsample of stars is: [Fe/H]=$-1.27$
($\sigma$=0.35, 29 stars), in good agreement with the average values derived
from the spectroscopic analyses, once differences between metallicity scales
are properly taken into account.

The star-to-star comparison between photometric and spectroscopic abundances is
shown in Figure~\ref{f:fig5b}. On average, the difference is $0.30\pm
0.07$~dex, with photometric abundances being larger as expected. However,
there are three stars (A10214, A26525, and B3-01408)
that have large residuals between
photometric and spectroscopic abundances. The light curve of one of these
stars (A10214) does not fully satisfy the completness and regularity criteria
for a reliable application of Jurcsik \& Kovacs (1996) method (the star has a
large D$_m$ value, see Di Fabrizio et al. 2004) and, for all these stars the
reddening $E(B-V)$\ derived from the Fourier decomposition of the light curve
is quite discrepant with respect to the average of the other stars. This
suggests that the Fourier analysis for these stars may be unreliable. If they
are dropped, the average difference between the two sets of metal abundances
is $0.28\pm 0.05$~dex, in reasonable agreement with the expected offset
between the Jurcsik (1995) and Harris (1996) metallicity scales. The r.m.s of
the residuals (0.24~dex), agrees well with observational errors of both
metallicity determinations (0.17~dex from spectroscopy; $\sim 0.2$~dex from
photometry).

\section{The luminosity - metallicity relation for RR Lyrae stars}

As discussed in the Introduction the metallicity [Fe/H] influences the
absolute magnitude of the RR Lyrae stars M$_{\rm V}$(RR) through a relation
generally assumed to be of linear form:  M$_{\rm V}$(RR)=$\alpha
\times$[Fe/H]+$\beta$. Rather large uncertainties exist on the slope $\alpha$
of this relation whose derivation was the ultimate goal of the present
spectroscopic study.

The LMC bar is the ideal place for defining the slope of the M$_{\rm
V}$(RR)-[Fe/H] relation since it contains a very numerous population of RR
Lyrae stars, spanning more than 1.5  dex in metallicity, and the variables are
all at the same distance from us, hence we do not have to worry about the
absolute magnitudes, i.e. about model assumptions. There still remains to
consider the intrinsic spread in the luminosities of the RR Lyrae stars due to
their evolution off the  ZAHB. To minimize this effect a fairly large number
of variables is  needed: we estimated that about 100 stars are required to
reduce the error bar in $\alpha$ down to better than 0.05 mag/dex.

We have combined the individual metal abundances of the RR Lyrae stars derived
from our spectroscopic study (see Column 13 of Table~\ref{t:basic}) with the
corresponding dereddened mean apparent V magnitudes derived from our
photometric work (see Column 4 of Table~\ref{t:basic}) to determine the slope
of the luminosity-metallicity relation for RR Lyrae stars with accuracy of
0.05 mag/dex. For 11 of the RR Lyrae variables we have analyzed
spectroscopically, the light curve is either incomplete, or with systematic
shifts between the 1999 and the 2001 photometry, or the star image appears to
be blended (see Columns 3-4 of Table~\ref{t:basic}), so that accurate values
of the average apparent magnitude could not be obtained for these stars. Our
luminosity-metallicity relation is thus based on the remaining 87 stars which
cover the metallicity range from [Fe/H]=$-0.79$ to $-2.12$. We have divided
the whole metallicity range into 5 bins, to have a reasonable number of
objects in each of them, and computed both the average metal abundance and the
average dereddened apparent luminosity $V_0$ and the corresponding rms errors
for each bin, these values are given in Table~\ref{t:tab6}.

\begin{table}
\caption{Metallicity bins used to calculate the luminosity-metallicity
relation of the LMC RR Lyrae stars}
\begin{tabular}{lcccc}
\hline
$[$Fe/H$]$~ bin & n. stars & $<[$Fe/H$]>$ & $<V_0>$ \\
\hline
~~~~~~~~~~$[$Fe/H$]<-1.8$ & ~9 & $-1.96\pm0.06$ & $18.99\pm 0.05$\\
$-1.8<[$Fe/H$]<-1.6$ & 18 & $-1.67\pm0.04$ & $19.01\pm 0.04$ \\
$-1.6<[$Fe/H$]<-1.4$ & 32 & $-1.49\pm0.03$ & $19.07\pm 0.03$ \\
$-1.4<[$Fe/H$]<-1.2$ & 21 & $-1.32\pm0.04$ & $19.09\pm 0.03$ \\
$-1.2<[$Fe/H$]$      & ~7 & $-0.97\pm0.07$ & $19.21\pm 0.06$ \\
\hline
\end{tabular}
\label{t:tab6}
\end{table}

From a least square fit of these average values weighted  by the errors in
both variables we derive the following relation between the apparent
luminosity and the metallicity of the RR Lyrae stars in our sample:
\begin{equation}
<V_0>=0.214 (\pm 0.047)({\rm [Fe/H]}+1.5)+ 19.064(\pm 0.017)
\end{equation}
where the error in the slope was evaluated via Monte Carlo simulations. The
relation is shown in Figure 19, where we distinguish between single and
double-mode RR Lyrae stars, indicated by open and filled squares respectively.
We note that the LMC RRd's are systematically offset to slightly higher
luminosities, thus suggesting that these stars may be more evolved than their
single-mode pulsator counterparts.

\begin{figure} 
\includegraphics[width=8.8cm]{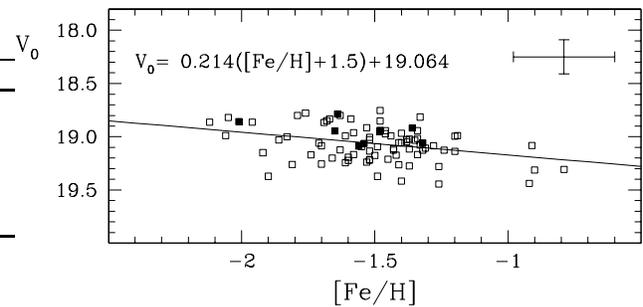}
\caption[]{Run of the dereddened mean magnitude of the RR Lyrae stars in our
two fields of the LMC as a function of the metallicity [Fe/H]. Filled symbols
are double-mode pulsators; open symbols ab- and c-type RR Lyrae stars.
}
\label{f:fig6}
\end{figure}
A complete discussion of the impact of the new luminosity-metallicity relation
has been given in Clementini et al. (2003a); here we only recall that the mild
slope we derive agrees very well both with the results by Rich et al. (2001)
based on the HB luminosity of 19 globular clusters in M31 ($\Delta {\rm
M_V(RR)}/\Delta$[Fe/H]= 0.22 mag/dex), and with the 0.20$\pm0.04$ slope found
by the Baade-Wesselink analysis of Milky Way field RR Lyrae stars (Fernley et
al. 1998). We also note that Figure~19 does not show clear evidence for a 
break in the slope around [Fe/H]=$-1.5$ as proposed by the theoretical models.
This seems to suggest that the luminosity-metallicity relation for HB stars is
universal, at least in M31, the LMC and the Milky Way. However, we caution the
reader that there are only 4 stars with metallicity larger than [Fe/H]=$-1$ in
our sample. A much larger number of stars with metallicities higher than
[Fe/H]=$-1.0$ would be necessary to definitely asses whether the M$_{\rm
V}$(RR)-[Fe/H] relation actually breaks at [Fe/H]=$-1.5$.

\section{The average metallicity of the short period Cepheids}

Our LMC sample includes 3 variables (star \#9604 and \#10320 in Field A, and
star \#5952 in field B) that have periods (0.29$<$ P $<$ 0.63 days) in the
range typical of the RR Lyrae stars but are from 0.5 to about 0.9 mag brighter
than the average luminosity of the HB in these fields (see left panels of
Figures 4 and 5 of Clementini et al. 2003a). The classification of these
variables posed some difficulty since, as discussed in Di Fabrizio et al.
(2004), their average luminosities, periods and amplitudes are consistent
either with those of the Anomalous Cepheids (ACs) commonly found in dwarf
Spheroidal galaxies (e.g. Pritzl et al. 2002, and references therein), or of
the low luminosity (LL) Cepheids (Clementini et al. 2003b) and of the short
period Classical Cepheids found in a number of dwarf Irregular galaxies (Smith
et al. 1992, Gallart et al. 1999, 2003, Dolphin et al. 2002).

Knowlewdge of the metallicy may help to classify these variables since there
is a limiting metallicity above which no ACs should be generated (Bono et al.
1997, Marconi et al. 2004)\footnote{This limit should be [Fe/H]$\sim -1.7$ for
variables around $\sim 1.3 M_\odot$ and [Fe/H]$\sim -2.3$ for variables around
$\sim 1.8 M_\odot$.}, and ACs are expected to have low metal abundances,
similar to or lower than the metallicity of the oldest populations in the host
system, while short period Classical Cepheids should have metallicities
similar to those of the Population I component in the system.

Two suspected ACs are known in $\omega$~Cen (Nemec et al. 1994, Kaluzny et al.
1997), a cluster known to span a wide range in metallicity with at least three
separate enrichment peaks (Norris, Freeman, \& Mighell 1996, Suntzeff \& Kraft
1996, Pancino et al. 2002) and suggested to possibly be the remnant of a
disrupted dwarf galaxy. One of them is among the stars in our sample (variable
\#73, see Table~6). The spectra of this star are shown in Figure~\ref{f:fig6},
those of the 3 LMC short period Cepheids are shown in Figure~\ref{f:fig7}.

\begin{figure} 
\includegraphics[width=8.8cm]{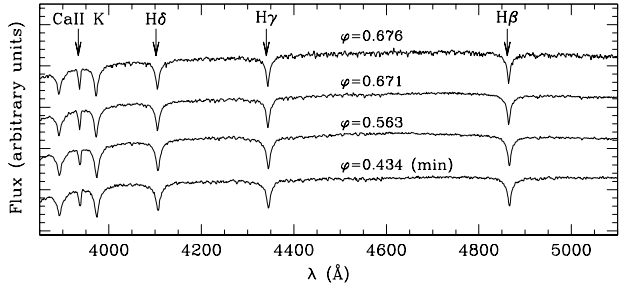}
\caption[]{Spectra of the variable star \#73, one of the two
suspected Anomalous Cepheids in $\omega$ Cen. 
For each spectrum we give the  phase along the 
pulsation cycle; the minimum light corresponds to the
phase interval 0.40$< \phi_{min} <0.52$ (bottom spectrum).
}
\label{f:fig6}
\end{figure}

\begin{figure} 
\includegraphics[width=8.8cm]{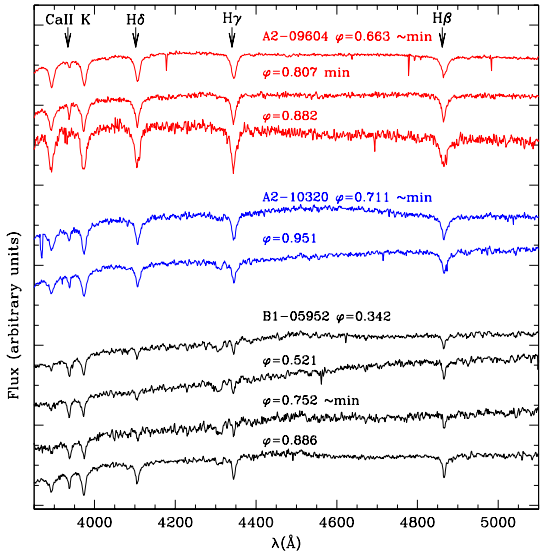}
\caption[]{Spectra of the 3 LMC SPC in our sample. For each spectrum we give
the corresponding phase along the 
pulsation cycle. The interval of minimum light is 0.77$< \phi_{min} <0.85$ for 
star A2-9604, $0.50 < \phi_{min} <0.68 $ for star A2-10320, and 
$0.77 < \phi_{min} <0.83 $ for star B1-5952.
}
\label{f:fig7}
\end{figure}

\begin{figure} 
\includegraphics[width=8.8cm]{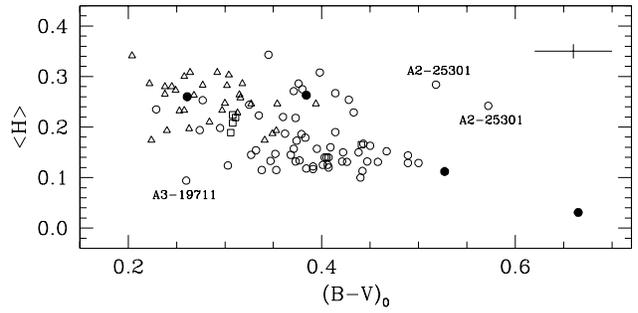}
\caption[]{Run of $<H>$ with the dereddened $B-V$ colour 
for the LMC variables observed around minimum light.
Different symbols are used for RRab (open circles), RRc (triangles), RRd 
(squares), and SPC (filled circles) variables. A few of the stars
shown in the figure have multiple observations.
}
\label{f:fig9}
\end{figure}

In the derivation of the spectroscopic abundances both the Anomalous Cepheid
in $\omega$ Cen (variable \#73), and the 3 short period Cepheids in the LMC
were analyzed as being normal horizontal branch stars. Indeed, the surface
gravity of these supra-HB stars should be similar to that of the RR Lyrae
variables, because they are brighter but also likely more massive. We then
expect that the metallicity calibration we use, based on normal horizontal
branch stars, should be roughly correct also for them. In order to check this
assumption in Figure~\ref{f:fig9} we show the run of the $<H>$ values versus
intrinsic (B$-$V)$_{0}$ colours for the LMC variables. These colours were
derived from the $B,V$ light curves (Di Fabrizio et al. 2004), dereddened
according to $E(B-V)_A$=0.116 and $E(B-V)_B$=0.086 (Clementini et al. 2003a),
and correspond to the colour at the phase of each individual spectrum. Only
data of stars observed around minimum light are displayed in
Figure~\ref{f:fig9}. Different symbols are used for the various types of
variables. Two of the RR Lyrae stars (A1-19711 and A2-25301) deviate from the
general trend, we suspect that the ephemerides and/or the reddening of these
two variables may be wrong. The short period Cepheids (filled symbols in
Figure~\ref{f:fig9}) follow the general $<H> - (B-V)_0$ relationship defined
by the RR Lyrae variables. The SPC spectra closer to minimum light fall on the
upper envelope at redder colours of this relation. This suggests that the
metallicity calibration we have adopted holds also for the SPCs. However, star
B1-05952 falls outside the range defined by the RR Lyrae variables, in a
totally extrapolated region of this relation, thus its metal abundance could
be slightly overestimated.
  
The [Fe/H] value we obtain for the $\omega$ Cen candidate Anomalous Cepheid
\#73 is [Fe/H]$=-1.50\pm 0.12$, in agreement with previous estimates (see
Table~6). This star appears to be marginally more metal-rich than the main
metal-poor component of $\omega$ Cen at [Fe/H]$\sim -1.6$ (Norris et al.
1996), but definitely metal-poorer than both the secondary metallicity peak at
[Fe/H]$\sim -1.2$ (Norris et al. 1996), and the metal-rich peak at [Fe/H]$\sim
-0.6$ (Pancino et al. 2002).

For the 3 short period Cepheids in our LMC sample we derive: [Fe/H]=$-1.96\pm
0.16$ for A2-09604, [Fe/H]=$-1.66\pm 0.09$ for A2-10320, and [Fe/H]=$-1.59:\pm
0.03$ for B1-05952, and a corresponding average value of [Fe/H]=$-1.74\pm
0.11$. Both individual and average metallicities of these three variables are
well below the average metal abundance of the RR Lyrae stars in our sample
([Fe/H]=$-1.48\pm 0.03$) thus suggesting that these three short period
Cepheids are more likely to be Anomalous Cepheids with masses $M \sim 1.3
M_\odot$ rather than the short period tail of the LMC Classical Cepheids.

It would be important to derive metallicities for more Anomalous Cepheids in
different environments. More extensive samples of Anomalous Cepheids with
adequate high resolution spectroscopic data would be clearly welcomed.

\begin{acknowledgements}
A special thanks goes to P. Montegriffo for the use of his software, to A.
Layden for providing us updated ephemerides for the NGC\,3201 variables in
advance of publication, and to A. Piersimoni for sending us finding charts for
the NGC\,3201 variables. We warmly thank the referee A. Walker for his 
constructive and helpful comments.
This research has made use of the SIMBAD data base,
operated at CDS, Strasbourg, France; it was funded by COFIN 2002028935 by
Ministero Universit\`a e Ricerca Scientifica, Italy
\end{acknowledgements}

\end{document}